\documentclass[notitlepage,nofootinbib,showpacs,preprintnumbers,superscriptaddress,amsmath,amssymb,11pt]{revtex4-2}

\usepackage{etoolbox}
\patchcmd{\section}
  {\centering}
  {\raggedright}
  {}
  {}
\patchcmd{\subsection}
  {\centering}
  {\raggedright}
  {}
  {}
\patchcmd{\subsubsection}
  {\centering}
  {\raggedright}
  {}
  {}

\makeatletter
\renewcommand{\p@subsection}{}
\renewcommand{\p@subsubsection}{}
\makeatother

\usepackage{graphicx}
\usepackage{dcolumn}
\usepackage{bm}
\usepackage{xcolor}
\usepackage[colorlinks=true,allcolors=blue]{hyperref}
\usepackage[normalem]{ulem}

\begin{document}


\title{Electron Sources for Accelerators}

\date{\today}
\author{D. Filippetto}
\affiliation{Berkeley National Laboratory, 1 Cyclotron Rd, Berkeley, CA 94720, USA}
\author{J. Grames}
\affiliation{Jefferson Lab., 12000 Jefferson Avenue, Newport News, VA 23606, USA}
\author{C. Hernandez-Garcia}
\affiliation{Jefferson Lab., 12000 Jefferson Avenue, Newport News, VA 23606, USA}
\author{S. Karkare}
\affiliation{ Arizona State University, 1151 S Forest Ave., Tempe, AZ 85281, USA}
\author{P. Piot}
\affiliation{Northern Illinois University, 1425 W Lincoln Hwy, DeKalb, IL 60115, USA}
\affiliation{Argonne National Laboratory, 9700 S. Cass Ave, Lemont, IL 60439, USA}
\author{J. Power}
\affiliation{Argonne National Laboratory, 9700 S. Cass Ave, Lemont, IL 60439, USA}
\author{Y. Sun}
\email{Corresponding author: yinesun@anl.gov}
\affiliation{Argonne National Laboratory, 9700 S. Cass Ave, Lemont, IL 60439, USA}
\author{E. Wang}
\affiliation{Brookhaven National Laboratory, 6 Lawrence Drive, Upton, NY 11973, USA}

\begin{abstract}
Electron sources are essential to an array of electron accelerator supporting research in high-energy physics and beyond. This report summarizes the``Snowmass 2021 Electron Source Workshop"~\cite{snowmasselectron21} which reviewed the current state-of-the art research and identified some possible research directions.
\end{abstract}

\maketitle

\section{Introduction}
Electron sources are critical components of accelerators used for high energy physics, nuclear physics, and light-sources. In conventional accelerators for particle and nuclear physics, the electron sources deliver bright, highly spin-polarized, high-charge electron beams, or ``main bunches'', which are ultimately employed as probes for physics. In addition to the main bunches, advanced accelerators based on beam-driven schemes also require electron sources capable of producing high-charge unpolarized ``drive'' bunches necessary to establishing large electromagnetic wakes in beam-driven plasma wakefield accelerator (PWFA) and structure-based wakefield accelerators (SWFA). Fututher more, high average current, unpolarized electron beams with low transverse emittance are used to ``cool'' other ``warm'' charged particle beams with larger thermal motion and transverse momentum, and thus increase the brightness of the particle beams for their applications in accelerators and colliders.  Electron sources also plays an important role in accelerator-based light sources, where electron beams with extremely small transverse emittance are often required.  In this White Paper, we outline the electron source requirements and the challenges in meeting these requirements for various components of electron sources, and discuss the directions for future research and development in different areas. 

\section{Electron Source Requirements}
Future electron-positron colliders such as CLIC, ILC, FCC-ee, CEPC etc are all designed to have high luminosity which implies high average current and small beam emittance beams. For a linear collider without damping rings, a flat beam with small vertical emittance will help to improve the luminosity as well. In addition, highly polarized electron sources are often required, which can be generated from a strained GaAs photocathode driven by a spin-polarized laser.  
\subsection{Bright-bunch generation}
Attaining high luminosity in a collider is contingent on the production of bright bunches. We identify two complementary research areas listed below.
\paragraph{Photocathodes and RF-gun selection:} The electron bunch would benefit from low-MTE photocathode such as the one currently investigated by other groups~\cite{hauri-2010a,droubay-2014a} while high QE is not critical given the low charges (2-3 orders of magnitude smaller than for the drive bunch). An additional challenge associated with the main-bunch production regards the requirement for spin-polarization often realized using NEA GaAs photocathodes, which is predominated utilized in a DC gun environment. The ultra-high-vacuum (UHV) required for GaAs photocathode has prevented their prolonged operation in RF guns~\cite{aleksandrov-2008a}. However, SRF guns could provide a path toward the reliable operation of spin-polarized photocathodes due to their excellent UHV levels sustained via cryo pumping. Ultimately, the 6D phase-space volume attainable in state-of-the-art photoinjectors combined with emittance-repartioning techniques could alleviate the need for an electron damping ring. 
\paragraph{Bunch shaping and beam manipulations:} Further improvements in the efficiency of an SWFA- and PWFA-based accelerators could be realized by shaping the main bunch current profile to load the wakefield generated by the drive bunch and significantly enhance the overall efficiency~\cite{tzoufras-2009a}. Controlling the bunch temporal distribution relies on similar laser-shaping techniques than the ones described above. Finally, the final beam distribution at the IP may be dominated by physics requirements (asymmetric “flat” beams~\cite{chen-1993a} or ultrashort bunches~\cite{yakimenko-2019a} for bremsstrahlung mitigation). These requirements call for small 6D phase-space volumes. Given the low bunch charge and shorter laser pulse involved in the emission process, a path to enhanced brightness are high-frequency RF guns capable of sustaining 100’s MV/m accelerating fields.

\subsection{High-charge drive-bunch generation}
In SWFA and PWFA-based colliders, the drive bunch exciting the wakefields is a high-charge electron beam with transverse emittance sufficient for transport through the structure(s). The formation of high-charge bunches shares many similarities with the main-bunch generation albeit at a much higher charge (but lower brightness). Producing such electron bunches relies on L-band photoinjectors and puts stringent demands on the associated photocathode and laser system especially when a high-repetition-rate is desired. Below we advocated three complementary paths for drive bunch research.
\paragraph{Laser and photocathode} To keep the photocathode-laser energy to sub-mJ level, the development of photocathodes with high-quantum-efficient (QE) and long lifetimes [O(week)] are needed. Also, photocathodes operating at wavelengths near available laser-media lasing range [typically in the infrared (IR) region of the spectrum] would be desirable to improve laser-to-electron wall-plug efficiency (by reducing the number of laser frequency conversion stages). Currently, Alkali-antimonide photocathodes are the most suited for high-charge bunches generation: CsK2Sb operates with ~530-nm laser pulses and has an acceptable mean transverse energy (MTE) of ~200 meV~\cite{bazarov-2011a}. Finally, improving the photocathode QE would be beneficial to further reduce the power requirement on the laser system and could be accomplished by, e.g., engineering the photocathode topology to enable plasmonic effects as demonstrated with metals~\cite{li-2013a}.
\paragraph{Space-charge mitigation:} The beam-dynamics associated with high-charge bunches is space-charge dominated during the bunch-formation process, initial acceleration, and transport to relativistic energy). Mitigating the impact of space-charge effects is critical to the generation of undisrupted beam distributions with reduced nonlinearities and improved emittances. Altering the photocathode topology was recently proposed as a way to alleviate the space-charge limit during the photoemission process~\cite{dowell-2019a} but a practical implementation and experimental validation remain unexplored. Likewise, shaping the laser distribution to linearize the space-charge fields offers a promising path to forming low-emittance high-charge bunches. The demonstrated blow-out regime~\cite{musumeci-2008a,piot-2013a} sets a stringent condition on the source size which ultimately limits the transverse emittance and alternative methods proposed to directly generate uniformly-filled 3D electron bunch by proper shaping of the laser distribution~\cite{limborg-2016a,li-2008a,kuzmin-2020a} should be experimentally tested. Likewise, laser shaping may also be adapted to program the laser temporal distribution to ultimately produce shaped drive bunches required for efficient beam-driven acceleration (improvement of transformer ratio (TR)~\cite{bane-1985a}); see example in Ref.~\cite{tan-2021a,xu-2019a}. Some of these laser-shaping techniques require precise control over the laser-pulse spectrum and would also benefit from photocathode operating close to the fundamental wavelength of the laser system.

\paragraph{RF-gun design \& beam dynamics:} The L-band RF guns typically employed to generate high-charge bunches provide a limited accelerating field on the cathode [typically {\cal O}(50 MV/m)] with 3-5-MeV final energy at the RF-gun exit. RF guns operating at higher frequencies could provide higher-field to mitigate collective effects but the smaller RF wavelength limit the bunch duration. Lower-frequency (VHF) RF-gun could offer a viable path owing to the long RF wavelength as long as the bunch exit energy remains high enough. For instance, numerical simulations indicate that a 200-MHz SRF gun with a 40-MV/m peak field~\cite{PhysRevLett.124.244801} is capable of producing a preshaped 10-nC electron bunch with parameters suitable to ultimately support an enhanced TR of $\sim 5$~\cite{tan-2021a} in an SWFA. These tradeoffs between RF-gun frequency choice and bunch quality should be studied quantitatively via precise numerical algorithms properly accounting for space charge effects during the emission process. The validity of the quasi-static models often employed to simulate the beam dynamics in photoinjectors should be investigated and validated against first-principle-physics algorithms (e.g. FDTD PIC algorithms). 

\section{Cathodes}
Photocathodes have very stringent and often conflicting requirements on the quantum efficiency (QE), the mean transverse energy (MTE) or the intrinsic emittance, robustness or operational lifetime, quick response time and spin-polarization \cite{musumeci2018advances}. For example, reduction in MTE can be obtained by operating at laser wavelengths closer to the threshold, however, this typically comes at the cost of the quantum efficiency. Due to such trade-offs there is no particular photocathode that suits all applications and cathode materials have to be chosen based on the needs of the specific application. The choice of the photocathode material is very closely related to the choice of the drive laser wavelength and the kind of electron gun used to extract electrons and in several cases also the design of the entire electron injector. For example, electron beams to be used as probes for high energy physics applications often need to be highly polarized with polarization exceeding 85\% \cite{Maruyama1992}. The only existing technology to obtain such highly polarized electrons is requires GaAs based photocathodes activated to negative electron affinity (NEA) operated at a specific wavelength close to the band-gap to optimize the QE and spin polarization. These cathodes are extremely sensitive to vacuum conditions and have only been successfully operated in DC guns for extended periods of time. Thus research on photocathode materials is very strongly linked to the developments in electron guns and laser technologies. This is true for all high average current, high charge, high peak brightness and high spin-polarization applications.

Many particle and nuclear physics experiments demand photocathodes with emitted electron spin-polarization exceeding 85-90\% along with a high QE in the single-digit-percent range to generate the probe beam \cite{Liu2016}. This level of performance has become the norm. Easily available $p$-doped GaAs wafers activated to negative electron affinity can provide polarization only as high as 35\% along with the required high QE. The higher polarization is typically obtained by growing strained GaAs and GaAsP hetero-structures which cause the splitting of the valence energy bands with different spins resulting in the very high polarization. However, due to the small market demand for such hetero-structures, a reliable source for obtaining such high-polarization cathodes no longer exists. This is likely to cause severe operational issues for existing and planned accelerator based particle and nuclear physics experiments. Thus developing a steady, reliable source of this existing hetero-structure based technology in the near-term is of utmost importance.

The second most critical issue is that of robustness and operational lifetime. These GaAs based cathodes are extremely sensitive to vacuum conditions and can only be operated in UHV DC high voltage guns with relatively low cathode fields at modest beam currents of $<$1 mA \cite{dunham2013record}. The ability to produce $>$100 mA of highly spin-polarized electron beam current from the cathode along with the robustness to operate in higher accelerating gradient SRF or RF guns to maximize brightness will enable novel designs of particle colliders which can achieve higher luminosity and are cheaper. The ERL-based design of the EIC where higher average current from the cathode is essential and the damping ring free design of the ILC where higher peak brightness from the cathode is essential are two examples of this. Thus developing robust sources of high QE, spin polarized electrons is a critical research direction for photocathodes. Lower MTE for spin-polarized sources will also be useful for increased luminosity.

Photocathodes producing unpolarized electrons are important not only as electron probes for experiments that don't demand spin-polarization, but are also useful for producing cold electron beams for hardon cooling and producing wakefields for advanced wakefield-based acceleration techniques.  Unpolarized electron beams used as electron probes and for hadron cooling applications are required to have average currents exceeding 100 mA.  Due to practical limitations on the achievable laser power cathodes required for high average current operation must have a QE exceeding 1\% in the visible wavelengths. Smaller intrinsic emittance is also beneficial. Alkali-antimonides (Cs$_3$Sb, K$_2$CsSb and Na$_2$KSb) have emerged as a strong contenders for such applications and have demonstrated average currents as high as 65 mA for several hours in a DC gun without significant QE degradation \cite{dunham2013record}. Despite their relative robustness to vacuum conditions compared to NEA-GaAs, these cathodes are still very sensitive to any oxidizing species and have demonstrated sustained operation only in DC and SRF guns. Ion-back bombardment and laser heating of the cathode are major causes of degradation at higher operating currents. Higher QE cathodes could alleviate the need for higher laser power addressing the latter. Effects of ion back-bombardment can be alleviated through use of more robust cathode materials and/or appropriate gun and injector designs to minimize it.  

Unpolarized electron bunches needed for exciting wakefields in advanced accelerator schemes require several nA of bunch charges, however, they do not require very large average currents due to lower repetition rates. This allows use of high QE, UV photocathodes like Cs$_2$Te \cite{Filippetto2015}. Despite the fact that Cs$_2$Te thin films require UHV, they are quite robust to vacuum conditions and have demonstrated successful operation in RF guns for extended periods of time of several months to over a year.

Research into developing robust high QE visible cathodes capable of operating in adverse vacuum environments, ideally air-stable, is critical for production of electron beams with average currents exceeding 100s of mA and beneficial for high bunch charge generation.   

Finally, beyond particle and nuclear physics applications, X-ray Free Electron Lasers (XFELs) and Ultrafast Electron Diffraction (UED) and Microscopy (UEM) require the brightest possible electron bunches and hence the smallest possible intrinsic emittace and operation of cathodes under the largest possible accelerating gradients~\cite{musumeci2018advances}. Brighter electron bunches will result in higher lasing photon energies and higher pulse energies from XFELs and also enable the development of compact XFELs. Higher brightness will increase the k-space resolution in UED experiments allowing studies of small crystals with larger lattice sizes and enable higher spatio-temporal resolution for UEM. Minimizing MTE requires tuning the incident laser wavelength very close to threshold, using materials with band-structure optimized for lower MTE, using cathodes with high QE close to threshold to minimize effects of non-linear photoemission and minimizing surface non-uniformities of physical roughness and work-function variations.

Research towards future cathode technologies for all the above applications calls for a collaborative theoretical and experimental effort between the fields of materials science, condensed matter physics and accelerator physics. Materials diagnostics techniques like Reflective High Energy Electron Diffraction, X-ray Absorption Spectroscopy, X-ray Reflectivity, X-ray Photoemission Spectroscopy, Auger Electron spectroscopy, Atomic Force Microscopy, optical characterization and several others can be used to characterize and control the growth of high QE thin film cathodes like alkali-antimonides and Cs$_2$Te. Efforts to develop epitaxial heterostructures of these materials can result in increased QE. Development of protective overlayers for activating GaAs to NEA and protecting vacuum sensitive alkali-antimonide cathodes also presents a promising area of future development. Exploration of other III-V materials especially in the III-nitrides family could result in more robust spin-polarized electron emitters. 
Theoretical and ab-initio modeling of photoemisison can allow virtual investigation of 10s of thousands of potential photocathode materials in a high throughput manner through the use of advanced materials databases and high performance computing. This can act as an efficient way of down selecting materials for detailed experimental investigations \cite{Antoniuk2021,chubenko2021monte}. Finally, testing the performance of novel photocathode candidate materials in electron guns under realistic operating conditions is critical. Development of cathode testing facilities is critical to achieve this.

\section{Electron guns for High Energy Physics Applications}
\subsection{Physics requirements}

The extreme beam parameters proposed for advanced collider concepts pose very stringent requirements on the electron source. This is particularly true for advanced linear collider, where the lower average current is balanced by much smaller beam sizes at the interaction point, and the beam-beam effects mitigated by compressing the pulses down below the picosecond. Here the initial beam 6D brightness, and the maximum average current achievable determine the achievable collider luminosity. 
In copper-based design (such as CCC~\cite{breidenbach_ccc_nodate} and CLIC~\cite{CLIC}) the driving RF pulse length is limited to $<1$-to-few \textmu{s}, at repetition rates of 50-120 Hz. A single RF pulse accelerates hundreds of bunches spaced by a few nanoseconds. Collider designs based on plasma wakefield acceleration techniques plan for a very different pulse format, with equidistant pulses at tens of kHz repetition rates~\cite{nghiem_toward_2020}. An electron source for such accelerators would therefore need to be able to provide high brightness electron beams capable of being efficiently squeezed into the micrometer-size accelerating region, at high repetition rates. 
When used for hadron cooling, the requirement for average current generated by the electron source becomes even larger($>$100 mA)~\cite{EIC}.

High average current and low emittance requirements inevitably lead to the use of semiconductor photo-cathodes, capable of providing higher quantum efficiency and lower MTE with respect to metal cathodes. The presence of such materials in the electron gun poses strict requirements on the absolute and partial vacuum levels in the accelerating gap. Indeed the high chemical reactivity of these photo-cathodes severely limits their lifetime, and requires the installation of load-lock systems for in-vacuum exchange of photocathode plugs, adding mechanical complexity to the electron gun region and to the electromagnetic design of the electron gun. 
Generation of polarized electron beams, included in most schemes of electron-based colliders to further enhance the collider luminosity and improve the experiment signal-to-noise, would require even more extreme vacuum performance, which can only be presently met in DC guns, as we will see below.  

Compatibility with strong magnetic fields at the cathode plane could also lead to an advantage in the design of electron guns for future colliders. Indeed, large beam transverse ratios are desired at the IP to minimize the luminosity degradation from beamstrahlung effects. The ratio between horizontal and vertical emittance required to obtain this asymmetry at the IP is obtained via the use of large damping rings. On the other hand, the same effect could be achieved by generation of "flat beams", via the use of strong magnetic fields at the cathode. The strong coupling between transverse planes can be used to produce a final beam with a large emittance ratio between the two transverse planes~\cite{piot_photoinjector_2006}. 

In the following we provide a brief summary of the main electron gun technologies, highlighting the main advantages and challenges of relevance for collider applications.

\subsection{Electron gun technologies}
\subsubsection{DC guns}
\paragraph{Polarized electron beam sources}
DC photoguns are the only currently available technology that has demonstrated – in the presence of accelerating gradients – the  extreme high vacuum conditions (10$^{-12}$ Torr) required by delicate strain super lattice (SSL) GaAs-based photocathodes. Today, only the photoguns at CEBAF and Mainz still regularly produce spin-polarized beams with $\sim$10 pC bunch charge in $\sim$50 ps-long FWHM pulses at 130 and 100 keV, respectively ~\cite{grames2018milliampere,friederichstatus}.  These voltages are sufficient to satisfy the electron beam requirements of the accelerators they serve, with no observable field emission hindering photocathode lifetime. However, meeting the stringent requirements of the EIC spin-polarized electron beam (5-7 nC per bunch, 4.5 A peak in a bunch train with 8 bunches per second) entails operating above 280 kV to manage space charge forces ~\cite{wang2022high}.  Higher operating voltage and extreme high vacuum conditions are competing factors. The best possible vacuum conditions in the accelerating gap are necessary to maximize the photocathode charge lifetime, which is limited by ion back-bombardment ~\cite{suleiman2018high}. 
Extending the photocathode lifetime by repelling ions incoming from the downstream beam line with a biased anode has proven very successful at CEBAF ~\cite{yoskowitz2021improving} and more recently at BNL ~\cite{wang2022high}. These two polarized electron beam sources represent the state of the art in DC photoguns for production of spin-polarized electron beams. Both employ conical ceramic insulators that extend from the top into the vacuum chamber (inverted insulator design) reducing overall gas load, in contrast with photoguns that employ large bore cylindrical insulators ~\cite{dunham2013record,kayran2018lerec, nishimori2013generation,hernandez2005high}.
The CEBAF photogun operating at 130 kV has generated $\sim$0.2 mA CW of highly polarized (85\%) beam for  nuclear physics experiments in more than a decade of operations, with a charge lifetime of 200-400 Coulombs ~\cite{suleiman2018high}. Extending this lifetime to 1000 Coulombs and $\sim$1 mA CW for an envisioned spin-polarized positron beam program at JLab presents a technological challenge ~\cite{abbott2016production}, but it is clear that managing ion back-bombardment is an essential factor in the photogun design. 
Operating at 300 kV DC without measurable field emission, and a novel cathode active cooling system integrated into the high voltage cable connector, the BNL polarized electron source represents the pinnacle of design incorporating many features for ion-back bombardment management such as: a biased anode, large anode aperture for clean beam transmission, large cathode electrode to reduce gradient and allowing large laser illumination area for distributing ion damage, and vacuum conditions approaching those in the CEBAF photogun ~\cite{grames2018milliampere,wang2022high}. The BNL photogun has surpassed the EIC beam requirements using bulk GaAs photocathodes (gap voltage, peak and average current, and bunch charge) in a test beam line at Stoney Brook University. Yet to be demonstrated using SSL GaAs photocathodes is generation of $>$85\% spin polarization electron beams, spin angle manipulation/measurements, and demonstration of $>$14 days photocathode lifetime at 56 nA average current. Preliminary results are promising after observing no measurable QE decay at $\sim$1000X that beam current from bulk GaAs photocathodes with biased anode ~\cite{wang2022high}. At 20 mA average current, the LHeC proposed polarized source lays beyond the state of the art. 

\paragraph{Unpolarized electron beam sources}
DC photoguns are also a mature technology for generating tens of mA CW un-polarized beams.  The maximum achievable CW current is limited by photocathode lifetime. The Cornell photogun operating at 250 kV demonstrated 65 mA CW from a CsK2Sb photocathode with $\sim$1 day lifetime ~\cite{dunham2013record}. A copy of the Cornell photogun developed at BNL for bunched beam cooling at RHIC demonstrated sustained 20 mA CW beam delivery operating at 375 kV with $\sim$6 days photocathode lifetime ~\cite{gu2020stable}. These photoguns represent the state-of-the-art in high CW beam current generation, requiring higher voltage than polarized sources to manage space charge. Based on large bore cylindrical segmented insulator rings with interior metal shields, this type of photoguns are capable of higher bias voltage than the inverted insulator designs, but achieving the extreme vacuum conditions typical of inverted insulator photogun is very challenging due to their larger surface area. The robustness of multi-alkali photocathodes allows for less stringent vacuum conditions, but UHV conditions during high current beam operations is still essential to minimize photocathode ion-back bombardment. Demonstrating the EIC requirements (100 mA, 1.5 nC bunch charge, 1.5 mm-mrad emittance and $>$3 days photocathode lifetime) is still beyond the state-of-the-art large-bore insulator DC photoguns.

\subsubsection{Normal-Conducting RF guns}

Radiofrequency electron guns based on normal-conducting technology have reached a high level of design maturity, and are today used as drivers in electron injectors for user facilities. Their development can be separated by the final operational goal, in pulsed and CW systems. 
The scaling of the breakdown field with RF frequency (Kilpatrick criterion~\cite{kilpatrick_criterion_1957}), makes high frequency (GHz) RF guns suitable for high gradient operations, while the scaling of the power density dissipated on the cavity wall with the frequency, favors lower frequency, in the VHF region, for high repetition rate operations. 

\paragraph{\textbf{Pulsed RF guns:}} Pulsed electron gun using RF frequency in the GHz range operate routinely with accelerating fields larger than $\sim 100$~MV/m and output energies in the multi MeV region, serving applications like FELs, inverse Compton scattering, and generating drive bunches for wakefield accelerators~\cite{dowell_rao_book}. The use of such technology allows the generation of low emittance, with nC-scale charge beams and picosecond-long pulse lengths, with a larger beam density thanks to order-of-magnitude increase in accelerating field respect to DC guns.
 
Most of the R\&D on this topic has been directed towards increasing the accelerating gradient, which would have a direct impact on the beam brightness.  
The breakdown limits for a specific cavity are associated RF frequency, cavity vacuum levels, surface roughness driving field emission~\cite{norem_triggers_2005}, and with the RF pulse duration, which also affects the total integrated dark current generated from the cavity walls and edges. Most of the electron guns are designed as standing wave cavities. Here the minimum RF pulse duration is set by the cavity filling time~$\tau_{RF}$, ranging from a few to few hundreds $\mu$s. Some RF designs utilize overcoupling to shorten $\tau_{RF}$, at the expenses of reduced power delivery to the cavity due to the consequent impedance mismatch. Typical cavities require multi-MW peak RF power to establish 50-100 MV/m acceleration fields. Typical  gun operations are limited to around 1000 Hz, due to the RF-induced heat load on the structure surfaces. 

More recently electron beam acceleration from a high-gradient electron gun powered by ns-long RF pulse has been demonstrated at the AWA complex~\cite{AWA_TW}, at Argonne National Laboratory. In this experiment the short RF pulses were generated by extracting power from a high-charge electron beam transported through the main AWA accelerator. The extracted 10~ns-long pulses at 11.7 GHz were then coupled into an X-band gun, establishing electric field on the cathode close to $\sim 400$~MV/m. This result demonstrates the potential of powering cavities with nanoseconds RF pulse to achieved gradients beyond current limitations.  

Another very promising recent development in the context of pulsed RF electron guns is the proposal of operating the structure at cryogenic temperatures, as copper at cryogenic temperatures has significantly lower resistivity loss and can withstand much higher surface fields~\cite{Cahill_cryoxband_18, Cahill_rfloss_18}. Such solution would also allow to increase at the same time the repetition rate of the source and its maximum accelerating field, pushing the limits of both peak and average beam brightness ~\cite{Rosenzweig_cryogun_19}. 

Finally, Load-lock systems for advanced photocathodes are now starting to be implemented in high-field pulsed RF guns, which will allow testing of semiconductor cathodes is such environments, possibly leading to a further decrease of beam emittance.

\paragraph{\textbf{CW RF guns:}}

In order to increase the repetition rate and the average current further into the MHz range, electron guns need to be operated in CW mode. In order to keep the power density dissipated on the cavity wall to a manageable level ($<$100W/cm$^2$), lower frequency designs have been explored. For a given energy gain, the power density is a steep function of the RF frequency ($f_{RF}^{5/2}$ 
), quantifying the tradeoff between higher fields (Breakdown limit, towards higher frequency) and higher average currents, (power density limit, towards lower frequencies).
Continuous-wave room-temperature normal-conducting RF guns have been developed in the context of high rep-rate X-FELs. For example, the VHF gun~\cite{APEXgun}, is designed with a reentrant-cavity profile at 186 MHz, enhancing the field in the accelerating gap, and obtaining fields in eccess  of 20 MV/m with a kinetic energy of up to 800 keV. The continuous input power needed to obtain such fields is of the order of 100~kW, with about 2 Joules of stored energy in the cavity. With this parameters, the power density dissipated on the walls is maintained below 25 W/cm$^2$ around the cavity. The gun is presently been used in high rep-rate UED beamlines ~\cite{Filippetto2016}, and to drive the LCLS-II XFEL ~\cite{lclsiigun2014}. 
Intense effort on the beam dynamics for the the VHF electron source (driven by LCLS-II needs) has led to a design of a low energy beam transport and compression section deviating substantially from the low-energy design for GHz guns. Taking advantage of the lower frequency and, therefore, of the larger temporal acceptance, a typical VHF gun producing high-charge beams operates in the so-called “cigar” regime, with a small aspect ratio A=R/L (defined at the cathode). Such regime has been shown to increase the maximum current density, and therefore the 5D transverse brightness, extractable from the cathode~\cite{MaxCurrent}. This is the main reason why typical transverse emittance performance of a VHF gun are comparable with the high-frequency guns. Such operating mode comes at the expenses of longitudinal emittance and beam current. In order to increase the peak current at the exit of the injector (tens of Amperes at 100 MeV) a high frequency RF bunching cavity is installed in the low energy region, close to the electron gun, with a second solenoid right before entering the linac sections. While this setup has been shown (in simulations) to be suited for driving an X-FEL, great care must be taken in managing non-linearities in the longitudinal phase space, which could potentially drastically limit the final peak current. 
One of the main advantages of Normal-conducting CW RF guns relates to the complexity of including systems for in-vacuum exchange of high quantum efficiency cathodes (load-lock systems). As of today, such systems have been routinely used to swap semiconductor cathodes including Cs$_2$Te and CsK$_2$Sb~\cite{filippetto_cesium_2015}, with cathode exchange times of 20 minutes. The vacuum in the load lock system is of the order of 1e-11, similar to the one in the cavity plenum in absence of RF field. CW operations increase the temperature of the cavity walls, leading to a steady-state vacuum performance about one order of magnitude larger. Lifetimes of Cs$_2$Te in this configuration has been shown to be compatible with user facility requirements. On the other hand, such electron guns have only been operated at maximum average currents of 1 mA, and it remains to be seen whether the cathode would suffer damage from back-bombardment during operations at higher currents, given the vacuum performance at least one order of magnitude higher than in DC or SRF guns. It is worth noting that the vacuum level reported here have not been exploiting the maximum pumping speed possible, and vacuum performance could be improved easily by utilizing the many ports available to add additional vacuum pumps.
 As a final comment, the VHF gun design was developed as a robust and reliable source in the context of a user facility. As such the RF performance of the cavity are not pushed to its maximum. There is ongoing effort to further improve the acceleration field to beyond 30 MV/m, approaching the limit of the allowable surface heat density~\cite{qian_cwgun_2019, pitz_cwgun_2019, apexii2017}.

\subsubsection{Superconducting RF guns}

SRF guns have the capability to generate high current electron beams in continuous-wave (CW) mode due to their cryogenic temperature operation, which reduces RF-induced wall heating at full duty cycle compared to normal conducting RF guns. The operating vacuum level is comparable to that in state-of-the-art dc photoguns utilized for generation of spin-polarized electron beams from SSL GaAs-based photocathodes. SRF guns represent the ultimate choice for generation of  high-brightness polarized electron beams thanks to the much higher gradient at the photocathode compared to that in a DC guns. However, integrating a normal conducting photocathode into the cryogenically cooled SRF cavity represents significant thermal and RF isolation technical challenges. Another reason for concern is contaminating the cavity with particulates during cathode insertion and from the photocathode itself when subject to high gradient during operation. The past decade has seen tremendous advances in SRF guns technologies ~\cite{xiang2021review}, from proof of principle using Pb films on the back wall of the Nb cavity ~\cite{vogel2019status}, to the routine use of normal conducting metal and semiconductor photocathodes for user-based accelerators at HZDR ~\cite{teichert2021successful,teichert2014free}, and for the Coherent Electron Cooling (CeC) proof of principle project at BNL ~\cite{xin2016design}. These two SRF guns represent the state-of-the-art, demonstrating that both metal and semiconductor photocathodes (cold or warm) can be safely operated to routinely deliver beam. 
The HZDR gun is an elliptical 3-1/2 cell cavity, with the capability to accept a NC Mg or Cu photocathode plugs, or a Cs$_2$Te semiconductor photocathode plug with RF choke. The cavity operates at 1.3 GHz at 2~K while the photocathode temperature is kept at ~80 K. Delivering CW beam at 200 pC at 100 kHz ( 20~\textmu{A} CW), this is the only SRF gun in existence that has provided beam for user-based FEL at ELBE. This was first achieved with Mg cathode extracting 57~Coulombs in 1760~h of user beam time between 2019 and 2020, and later 2020-2022 with a Cs$_2$Te photocathode extracting 91 Coulombs for over 3000~h of user beam time.
The BNL SRF gun is an 112~MHz Quarter Wave Resonator (QWR) cavity operating at 4 K while the CsK$_2$Sb photocathode with RF choke remains at room temperature. The gun has been in routine operation since 2016 and has demonstrated up to 10 nC bunch charge in ns-long pulses due to relatively low gradient at the photocathode. The highest average current the gun can currently provide is ~ 150~\textmu{A} CW limited by available RF power ~\cite{Wang_2021sr}. 
To meet the stringent emittance requirements for the LCLS-II-HE injector ($\simeq 100$~nm at 100 pC), the proposed SRF gun is a QWR based on the WiFEL design ~]\cite{legg2012status} but with higher cathode gradient ~\cite{xiang2021review}.
Although managing RF heat loads, photocathode thermal contact, minimizing particulate generation during cathode insertion and operation, and suppressing multipacting are active R\&D efforts and critical technical aspects to resolve, tests at HZB BerlinPro, KEK, and PKU continue to show steady progress for integrating multi-alkali photocathodes capable of delivering high-average-current beams when illuminated with 530-nm laser pulses ~\cite{xiang2021review}.

\subsubsection{Advanced source concepts}

This topic is reviewed in depth in ~\cite{fuchs_snowmass_2022}, where a complete list of experimental work and achievements can be found.  Here we will just briefly summarize the main directions. 

As noted in the previous sections, the presence of large accelerating fields at the "photocathode" plane lead to large beam brightness. In plasma-driven electron sources, the photocathode is represented by a laser-ionized gas rather than by a solid surface.  

Demonstrated advanced electron sources include self-injection regime in LWFA, where a high intensity laser ionizes a gas and generated an ionic region with very strong longitudinal fields. Electrons expelled from the region are then trapped, injected back into the region and accelerated to ultra-relativistic energies. Many variants of this concept have been proposed and demonstrated in the last decade, providing improved control of beam quality and stability~\cite{fuchs_snowmass_2022}. 
Another very promising advanced electron source recently demonstrated stems for the concept of PWFA, where a high-density electron pulse is used to drive a plasma wave inside a gas. In order to generate electron beams, an independent high intensity laser pulse is used to ionize the gas and inject electrons in the wake left by the electron beam. 
As of today, many concepts have been proposed for generation of polarized and unpolarized electron beams, but all schemes tested experimentally focus on generation of unpolarized electron beams. Experiments demonstrated generation of very bright beams, showcasing the enormous potential of the technique. On the other hand, many of the outstanding parameters have been demonstrated separately rather than simultaneously, and their use as drivers for high energy physics colliders, although of great interest, is still far from becoming a reality. 

\section{Injectors}
The electron injector includes the gun and subsequent beamline that boosts the energy of the electron beam out of the space charge dominated regime.  Depending on the charge of the beam, the energy at the exit of the injector ranges from 10s to 100s of MeV. The design of the injector varies according to the electron gun technology (see Section IV. Electron Guns), but in all cases, the injector is designed to preserve  the 6D phase space volume generated by the gun and to manipulate the phase space to meet the specifications of the trailing linac. Injectors can be classified into two categories: the continuous wave (CW) injector and the pulsed injector.  Further, CW injectors are based on three type of electron guns: high voltage direct current (HVDC) gun, superconducting Radio Frequency (SRF) gun and CW normal conducting RF gun while pulsed injectors are typically based normal conducting RF guns. Small emittance can be achieved from the gun with a appropriate combination of laser shaping, high gradient on the cathode and small MTE cathode material.  The gun emittance can be further controlled with phase space manipulation.

\subsection{Laser shaping to control space charge in the injector}
In the injector, space charge is the main source of emittance growth and occurs when high charge bunches are still in the space charge dominated regime. This is the regime where the transport of the beam is dominated by the space charge force compared with the emittance pressure. ~\cite{power-2005} Therefore, the injector design must consider space charge.  Emittance compensation is used to handle the linear space charge forces and is done with a solenoid at the exit of the gun. However, non-linear space charge forces, both transverse and longitudinal, cannot be compensated for with the solenoid-based method. One way to handle the non-linear space charge forces is with laser shaping.  Laser shaping can be used to control that charge distribution of the electron bunch in order to minimize the nonlinear space charge. The bunch shape that eliminates all non-linear space charge forces is the uniformly filled ellipsoid.  IR laser shaping has been demonstrated with a spatial light modulator (SLM) shaper and 3D Volume chirp Bragg grating~\cite{Mironov:16}. This main challenge of this IR-based SLM method is converting the IR shape to UV for use on the Cs$_2$Te or metal photocathodes. Using green-light photocathodes such as multialkali can reduce the challenge in laser shaping due to a single stage of frequency doubling. Other distributions, such as the truncated Gaussian or quasi 3D shapes have already been generated and provide approximately linear space charge.

\subsection{Phase space manipulation in the injector}
A small 6D emittance is essential for increasing the luminosity of the lepton collider.  In addition,  to suppress beamstrahlung, current linear colliders designs require special phase space distributions. For example, the International Linac Collider (ILC) ~\cite{ILC-2022, ilc_tdr}, requires a \emph{flat beam} distribution where the vertical emittance is much smaller than the horizontal while advanced linear colliders require ultrashort bunches. The flat beam distribution has historically been generated with a damping ring, but studies to use phase space manipulation to eliminate the high cost of the damping ring are underway~\cite{xu-2022a}. This 6D phase space manipulation can be achieved by a combination of a round to flat beam transformer (RFBT) and an emittance exchanger (EEX). The RFBT is done in the injector by using a magnetized electron beam and round to flat converter, such as three skew quads~\cite{PhysRevSTAB.6.104002,PhysRevE.66.016503,PhysRevSTAB.9.031001}. The RFBT method can be extended to develop a positron injector, which has a much higher initial transverse emittance than electron injector or to electron cooling where the injector is matched to the hadron beam shape in the ring~\cite{wang2021accelerator}. EEX based beam manipulation is also carried out in the injector which enables a longitudinal-transverse emittance exchange scheme by placing a transverse deflector cavity in the middle of four dipoles staircase~\cite{PhysRevSTAB.5.084001}.

\subsection{Pulsed and CW injectors}
Low average current injectors are normally based on normal conducting pulsed RF guns, typically S-band or L-band. These guns offer several advantages over CW guns due to the ability to operate with high-gradient electric fields on the photocathode, order 100 MV/m. This high gradient allows the gun to directly generate a pancake shaped beams (i.e. short electron bunches with large transverse dimensions) with smallest emittance at moderately relativistic energies, ~4-8 MeV.  Applications include linac-based free electron laser (FEL), ultrafast electron diffraction, and wakefield accelerators. Since the bunch from the NCRF gun is relatively stiff and has an initial short bunch length it is directly injected into a booster linac to accelerate the beam out of the space charge dominated regime.  If further bunch compression is needed, it can be done after the injector section with a magnetic chicane compressor.\\
Most of the normal conducting pulsed RF gun chose the standing wave mode due to its higher $(R/Q)$. A traveling wave (TW) photogun can reduce the RF filling time, therefore possible to achieve higher repetition rate. RF guns with multiple cells allows the beam to become relativistic at exit of the gun where the space charge induced emittance growth is reduced~\cite{LUCAS2021165651}. \\

High average current injectors are normally based on CW electron guns. Applications for these injectors include high power FELs, energy recovery Linac (ERL) based collider and electron cooling.  The CW SRF gun can generate high current, high bunch charge, and high brightness electron beam. It advantages compared to other CW guns include operation with a high gradient at the cathode (which allows the optimum emission phase to approach the crest of the accelerating field with large gap voltage), excellent baseline vacuum due to the large surface of cryo pumping at 4K operation. Very high frequency (VHF), quarter-wave (QW) SRF guns are used in several projects including the BNL 113 MHz gun for Coherent electron cooling-X, 200 MHz WiFEL gun for LCLS-II R\&D, and ~200 MHz gun for LCLS-II operation~\cite{Wang_2021sr,PhysRevLett.124.244801}. Due to its low frequency, it usually generates a Cigar-shape beam that is then compressed in a bunching section to achieve high peak current and small emittance. The space charge compensation scheme is still needed to reduce the emittance. Therefore, a solenoid is placed at the exit of the gun while carefully avoiding generating a residual axial solenoidal field in the superconductor material. Using a high-temperature superconducting (HTS) solenoid in the cryomodule can effectively generate sufficient field to compensate the linear space charge. The challenges of using SRF gun is how to come over the multipacting zones, get it operation stably and avoid contamination from semiconductor cathode material. \\

The CW NCRF gun was considered to have poor vacuum and was not suitable for polarized photocathodes (e.g. GaAs) before. Recent development shows a VHF NCRF gun can generate MHz repetition rate electron beams using photocathode at the design energy as well~\cite{vhf_15}. Unlike pulsed RF guns, both CW SRF and CW NCRF guns generate significant dark current in each RF cycle when the cathode surface gradient is higher than the field emission threshold. Because the semiconductor photocathode typically has a low work function, compared to the metal photocathode, the average current of the dark current beam may be higher than the main beam.  This can cause an issue of the downstream beam diagnostics not being able to detect the signal over the noise. Solution to reduce the dark current includes: polishing the cathode substrate, using a collimator or a dark current kicker in the injector. Reducing the dark current in the injector remains one of the most important topics for using CW electron source~\cite{Wang_2021sr,PhysRevSTAB.18.013401,PhysRevSTAB.17.043401}. \\

\section{An outlook for future electron sources}
We have reviewed the state-of-art for the various components of electron injectors. Area for research and development on electron sources in the future can be summarized as the following: 
\subsection{Cathodes}
\begin{itemize}
    \item 	Highly-polarized electron sources are required as probes for high energy physics. Developing a steady, robust, reliable source of the hetero-structure based cathode such as GaAs etc. in the near-term is of utmost importance.  The cathode lifetime study is strongly linked to the development in electron guns and drive laser technology. 
    \item 	Unpolarized electron source with high quantum efficiency, small intrinsic emittance is another area for cathode development. Research towards future cathode technologies calls for a collaborative theoretical and experimental effort between the fields of materials science, condense matter physics and accelerator physics.
    \item 	Development of cathode testing facilities is critical in testing the performance of novel photocathode candidate materials in electron guns under realistic operating conditions.
\end{itemize}
\subsection{Guns}
DC guns have better vacuum which allows a better lifetime for polarized GaAs cathode. Highest average current under CW (continuous wave) operations were demonstrated with DC guns. RF guns can reach higher accelerating gradient which is essential for low beam emittance and high brightness. SCRF guns have comparable vacuum conditions with DC guns but can provide much higher accelerating gradient.
\begin{itemize}
\item 	DC guns: reaching for high operating voltage to mitigate space-charge forces, maintaining  high vacuum conditions to preserve the GaAs-based photocathode life time, and managing ion back-bombardment are important area for future R\&D.
\item NCRF guns: Pushing for high accelerating gradient in the direction of high RF frequency (GHz) guns. Copper structures operated in cryogenic temperatures have the potential for both higher gradient and repetition rate.  Higher average current is more manageable in lower frequency RF guns.
\item SCRF guns: normal conducting photocathode insertion into a SCRF gun is a challenging aspect in the SCRF gun technology. Collaborations with leading labs (such as HZDR  and BNL) should be encouraged for the development of SCRF guns in USA, such as the LCLS-II-HE injector SCRF gun injector project.
\end{itemize}

\subsection{Injectors}
The electron injector delivers beam up to $\sim 100$~MeV. The injectors can be either CW or pulsed. 
\begin{itemize}
\item Achieving high 6D phase-space brightness and manipulating the phase-space distribution of the electron beam to fits it applications should be invested. This includes the photo-injector drive-laser shaping, round-to-flat beam transformation, emittance exchange between transverse and longitudinal phase space, pulse train generation etc. 
\item Reducing dark current is one of the most important topics for CW electron injectors.
\end{itemize}
\subsection{Advanced concepts}
LWFA and PWFA based electron sources focused on the generation of unpolarized electrons have demonstrated impressive beam parameters but their application as a driver for HEP colliders are still far from becoming a reality.

%

\bibliography{esource,Eguns}

\begin{thebibliography}{78}%
\makeatletter
\providecommand \@ifxundefined [1]{%
 \@ifx{#1\undefined}
}%
\providecommand \@ifnum [1]{%
 \ifnum #1\expandafter \@firstoftwo
 \else \expandafter \@secondoftwo
 \fi
}%
\providecommand \@ifx [1]{%
 \ifx #1\expandafter \@firstoftwo
 \else \expandafter \@secondoftwo
 \fi
}%
\providecommand \natexlab [1]{#1}%
\providecommand \enquote  [1]{``#1''}%
\providecommand \bibnamefont  [1]{#1}%
\providecommand \bibfnamefont [1]{#1}%
\providecommand \citenamefont [1]{#1}%
\providecommand \href@noop [0]{\@secondoftwo}%
\providecommand \href [0]{\begingroup \@sanitize@url \@href}%
\providecommand \@href[1]{\@@startlink{#1}\@@href}%
\providecommand \@@href[1]{\endgroup#1\@@endlink}%
\providecommand \@sanitize@url [0]{\catcode `\\12\catcode `\$12\catcode
  `\&12\catcode `\#12\catcode `\^12\catcode `\_12\catcode `\%12\relax}%
\providecommand \@@startlink[1]{}%
\providecommand \@@endlink[0]{}%
\providecommand \url  [0]{\begingroup\@sanitize@url \@url }%
\providecommand \@url [1]{\endgroup\@href {#1}{\urlprefix }}%
\providecommand \urlprefix  [0]{URL }%
\providecommand \Eprint [0]{\href }%
\providecommand \doibase [0]{https://doi.org/}%
\providecommand \selectlanguage [0]{\@gobble}%
\providecommand \bibinfo  [0]{\@secondoftwo}%
\providecommand \bibfield  [0]{\@secondoftwo}%
\providecommand \translation [1]{[#1]}%
\providecommand \BibitemOpen [0]{}%
\providecommand \bibitemStop [0]{}%
\providecommand \bibitemNoStop [0]{.\EOS\space}%
\providecommand \EOS [0]{\spacefactor3000\relax}%
\providecommand \BibitemShut  [1]{\csname bibitem#1\endcsname}%
\let\auto@bib@innerbib\@empty
\bibitem [{\citenamefont {{Sun (Chair)}}()}]{snowmasselectron21}%
  \BibitemOpen
  \bibfield  {author} {\bibinfo {author} {\bibfnamefont {Y.}~\bibnamefont {{Sun
  (Chair)}}},\ }\href {https://indico.fnal.gov/event/46053/} {\bibinfo {title}
  {Snowmass2021 electron source workshop}},\ \bibinfo {note} {virtual workshop
  organized by Argonne National Laboratory, Feb 16 – 18, 2022}\BibitemShut
  {NoStop}%
\bibitem [{\citenamefont {Hauri}\ \emph {et~al.}(2010)\citenamefont {Hauri},
  \citenamefont {Ganter}, \citenamefont {Le~Pimpec}, \citenamefont {Trisorio},
  \citenamefont {Ruchert},\ and\ \citenamefont {Braun}}]{hauri-2010a}%
  \BibitemOpen
  \bibfield  {author} {\bibinfo {author} {\bibfnamefont {C.~P.}\ \bibnamefont
  {Hauri}}, \bibinfo {author} {\bibfnamefont {R.}~\bibnamefont {Ganter}},
  \bibinfo {author} {\bibfnamefont {F.}~\bibnamefont {Le~Pimpec}}, \bibinfo
  {author} {\bibfnamefont {A.}~\bibnamefont {Trisorio}}, \bibinfo {author}
  {\bibfnamefont {C.}~\bibnamefont {Ruchert}},\ and\ \bibinfo {author}
  {\bibfnamefont {H.~H.}\ \bibnamefont {Braun}},\ }\bibfield  {title} {\bibinfo
  {title} {Intrinsic emittance reduction of an electron beam from metal
  photocathodes},\ }\href {https://doi.org/10.1103/PhysRevLett.104.234802}
  {\bibfield  {journal} {\bibinfo  {journal} {Phys. Rev. Lett.}\ }\textbf
  {\bibinfo {volume} {104}},\ \bibinfo {pages} {234802} (\bibinfo {year}
  {2010})}\BibitemShut {NoStop}%
\bibitem [{\citenamefont {Droubay}\ \emph {et~al.}(2014)\citenamefont
  {Droubay}, \citenamefont {Chambers}, \citenamefont {Joly}, \citenamefont
  {Hess}, \citenamefont {N\'emeth}, \citenamefont {Harkay},\ and\ \citenamefont
  {Spentzouris}}]{droubay-2014a}%
  \BibitemOpen
  \bibfield  {author} {\bibinfo {author} {\bibfnamefont {T.~C.}\ \bibnamefont
  {Droubay}}, \bibinfo {author} {\bibfnamefont {S.~A.}\ \bibnamefont
  {Chambers}}, \bibinfo {author} {\bibfnamefont {A.~G.}\ \bibnamefont {Joly}},
  \bibinfo {author} {\bibfnamefont {W.~P.}\ \bibnamefont {Hess}}, \bibinfo
  {author} {\bibfnamefont {K.}~\bibnamefont {N\'emeth}}, \bibinfo {author}
  {\bibfnamefont {K.~C.}\ \bibnamefont {Harkay}},\ and\ \bibinfo {author}
  {\bibfnamefont {L.}~\bibnamefont {Spentzouris}},\ }\bibfield  {title}
  {\bibinfo {title} {Metal-insulator photocathode heterojunction for directed
  electron emission},\ }\href {https://doi.org/10.1103/PhysRevLett.112.067601}
  {\bibfield  {journal} {\bibinfo  {journal} {Phys. Rev. Lett.}\ }\textbf
  {\bibinfo {volume} {112}},\ \bibinfo {pages} {067601} (\bibinfo {year}
  {2014})}\BibitemShut {NoStop}%
\bibitem [{\citenamefont {Aleksandrov}\ \emph {et~al.}(2008)\citenamefont
  {Aleksandrov}, \citenamefont {Konstantinov}, \citenamefont {Logatchov},
  \citenamefont {Novokhatski},\ and\ \citenamefont {A.A}}]{aleksandrov-2008a}%
  \BibitemOpen
  \bibfield  {author} {\bibinfo {author} {\bibfnamefont {A.}~\bibnamefont
  {Aleksandrov}}, \bibinfo {author} {\bibfnamefont {E.}~\bibnamefont
  {Konstantinov}}, \bibinfo {author} {\bibfnamefont {P.}~\bibnamefont
  {Logatchov}}, \bibinfo {author} {\bibfnamefont {A.}~\bibnamefont
  {Novokhatski}},\ and\ \bibinfo {author} {\bibfnamefont {S.}~\bibnamefont
  {A.A}},\ }\bibfield  {title} {\bibinfo {title} {The accelerator design
  progress for eic strong hadron cooling},\ }in\ \href@noop {} {\emph {\bibinfo
  {booktitle} {Proceedings of the 2008 European Particle Accelerator Conference
  (EPAC08)}}}\ (\bibinfo {year} {2008})\ p.\ \bibinfo {pages}
  {1451}\BibitemShut {NoStop}%
\bibitem [{\citenamefont {Tzoufras}\ \emph {et~al.}(2009)\citenamefont
  {Tzoufras}, \citenamefont {Lu}, \citenamefont {Tsung}, \citenamefont {Huang},
  \citenamefont {Mori}, \citenamefont {Katsouleas}, \citenamefont {Vieira},
  \citenamefont {Fonseca},\ and\ \citenamefont {Silva}}]{tzoufras-2009a}%
  \BibitemOpen
  \bibfield  {author} {\bibinfo {author} {\bibfnamefont {M.}~\bibnamefont
  {Tzoufras}}, \bibinfo {author} {\bibfnamefont {W.}~\bibnamefont {Lu}},
  \bibinfo {author} {\bibfnamefont {F.~S.}\ \bibnamefont {Tsung}}, \bibinfo
  {author} {\bibfnamefont {C.}~\bibnamefont {Huang}}, \bibinfo {author}
  {\bibfnamefont {W.~B.}\ \bibnamefont {Mori}}, \bibinfo {author}
  {\bibfnamefont {T.}~\bibnamefont {Katsouleas}}, \bibinfo {author}
  {\bibfnamefont {J.}~\bibnamefont {Vieira}}, \bibinfo {author} {\bibfnamefont
  {R.~A.}\ \bibnamefont {Fonseca}},\ and\ \bibinfo {author} {\bibfnamefont
  {L.~O.}\ \bibnamefont {Silva}},\ }\bibfield  {title} {\bibinfo {title} {Beam
  loading by electrons in nonlinear plasma wakes},\ }\href
  {https://doi.org/10.1063/1.3118628} {\bibfield  {journal} {\bibinfo
  {journal} {Physics of Plasmas}\ }\textbf {\bibinfo {volume} {16}},\ \bibinfo
  {pages} {056705} (\bibinfo {year} {2009})},\ \Eprint
  {https://arxiv.org/abs/https://doi.org/10.1063/1.3118628}
  {https://doi.org/10.1063/1.3118628} \BibitemShut {NoStop}%
\bibitem [{\citenamefont {Chen}(1993)}]{chen-1993a}%
  \BibitemOpen
  \bibfield  {author} {\bibinfo {author} {\bibfnamefont {P.}~\bibnamefont
  {Chen}},\ }\bibfield  {title} {\bibinfo {title} {Disruption effects from the
  collision of quasi-flat beams},\ }in\ \href
  {https://doi.org/10.1109/PAC.1993.308793} {\emph {\bibinfo {booktitle}
  {Proceedings of International Conference on Particle Accelerators}}}\
  (\bibinfo {year} {1993})\ pp.\ \bibinfo {pages} {617--619 vol.1}\BibitemShut
  {NoStop}%
\bibitem [{\citenamefont {Yakimenko}\ \emph {et~al.}(2019)\citenamefont
  {Yakimenko}, \citenamefont {Meuren}, \citenamefont {Del~Gaudio},
  \citenamefont {Baumann}, \citenamefont {Fedotov}, \citenamefont {Fiuza},
  \citenamefont {Grismayer}, \citenamefont {Hogan}, \citenamefont {Pukhov},
  \citenamefont {Silva},\ and\ \citenamefont {White}}]{yakimenko-2019a}%
  \BibitemOpen
  \bibfield  {author} {\bibinfo {author} {\bibfnamefont {V.}~\bibnamefont
  {Yakimenko}}, \bibinfo {author} {\bibfnamefont {S.}~\bibnamefont {Meuren}},
  \bibinfo {author} {\bibfnamefont {F.}~\bibnamefont {Del~Gaudio}}, \bibinfo
  {author} {\bibfnamefont {C.}~\bibnamefont {Baumann}}, \bibinfo {author}
  {\bibfnamefont {A.}~\bibnamefont {Fedotov}}, \bibinfo {author} {\bibfnamefont
  {F.}~\bibnamefont {Fiuza}}, \bibinfo {author} {\bibfnamefont
  {T.}~\bibnamefont {Grismayer}}, \bibinfo {author} {\bibfnamefont {M.~J.}\
  \bibnamefont {Hogan}}, \bibinfo {author} {\bibfnamefont {A.}~\bibnamefont
  {Pukhov}}, \bibinfo {author} {\bibfnamefont {L.~O.}\ \bibnamefont {Silva}},\
  and\ \bibinfo {author} {\bibfnamefont {G.}~\bibnamefont {White}},\ }\bibfield
   {title} {\bibinfo {title} {Prospect of studying nonperturbative qed with
  beam-beam collisions},\ }\href
  {https://doi.org/10.1103/PhysRevLett.122.190404} {\bibfield  {journal}
  {\bibinfo  {journal} {Phys. Rev. Lett.}\ }\textbf {\bibinfo {volume} {122}},\
  \bibinfo {pages} {190404} (\bibinfo {year} {2019})}\BibitemShut {NoStop}%
\bibitem [{\citenamefont {Bazarov}\ \emph {et~al.}(2011)\citenamefont
  {Bazarov}, \citenamefont {Cultrera}, \citenamefont {Bartnik}, \citenamefont
  {Dunham}, \citenamefont {Karkare}, \citenamefont {Li}, \citenamefont {Liu},
  \citenamefont {Maxson},\ and\ \citenamefont {Roussel}}]{bazarov-2011a}%
  \BibitemOpen
  \bibfield  {author} {\bibinfo {author} {\bibfnamefont {I.}~\bibnamefont
  {Bazarov}}, \bibinfo {author} {\bibfnamefont {L.}~\bibnamefont {Cultrera}},
  \bibinfo {author} {\bibfnamefont {A.}~\bibnamefont {Bartnik}}, \bibinfo
  {author} {\bibfnamefont {B.}~\bibnamefont {Dunham}}, \bibinfo {author}
  {\bibfnamefont {S.}~\bibnamefont {Karkare}}, \bibinfo {author} {\bibfnamefont
  {Y.}~\bibnamefont {Li}}, \bibinfo {author} {\bibfnamefont {X.}~\bibnamefont
  {Liu}}, \bibinfo {author} {\bibfnamefont {J.}~\bibnamefont {Maxson}},\ and\
  \bibinfo {author} {\bibfnamefont {W.}~\bibnamefont {Roussel}},\ }\bibfield
  {title} {\bibinfo {title} {Thermal emittance measurements of a cesium
  potassium antimonide photocathode},\ }\href
  {https://doi.org/10.1063/1.3596450} {\bibfield  {journal} {\bibinfo
  {journal} {Applied Physics Letters}\ }\textbf {\bibinfo {volume} {98}},\
  \bibinfo {pages} {224101} (\bibinfo {year} {2011})},\ \Eprint
  {https://arxiv.org/abs/https://doi.org/10.1063/1.3596450}
  {https://doi.org/10.1063/1.3596450} \BibitemShut {NoStop}%
\bibitem [{\citenamefont {Li}\ \emph {et~al.}(2013)\citenamefont {Li},
  \citenamefont {To}, \citenamefont {Andonian}, \citenamefont {Feng},
  \citenamefont {Polyakov}, \citenamefont {Scoby}, \citenamefont {Thompson},
  \citenamefont {Wan}, \citenamefont {Padmore},\ and\ \citenamefont
  {Musumeci}}]{li-2013a}%
  \BibitemOpen
  \bibfield  {author} {\bibinfo {author} {\bibfnamefont {R.~K.}\ \bibnamefont
  {Li}}, \bibinfo {author} {\bibfnamefont {H.}~\bibnamefont {To}}, \bibinfo
  {author} {\bibfnamefont {G.}~\bibnamefont {Andonian}}, \bibinfo {author}
  {\bibfnamefont {J.}~\bibnamefont {Feng}}, \bibinfo {author} {\bibfnamefont
  {A.}~\bibnamefont {Polyakov}}, \bibinfo {author} {\bibfnamefont {C.~M.}\
  \bibnamefont {Scoby}}, \bibinfo {author} {\bibfnamefont {K.}~\bibnamefont
  {Thompson}}, \bibinfo {author} {\bibfnamefont {W.}~\bibnamefont {Wan}},
  \bibinfo {author} {\bibfnamefont {H.~A.}\ \bibnamefont {Padmore}},\ and\
  \bibinfo {author} {\bibfnamefont {P.}~\bibnamefont {Musumeci}},\ }\bibfield
  {title} {\bibinfo {title} {Surface-plasmon resonance-enhanced multiphoton
  emission of high-brightness electron beams from a nanostructured copper
  cathode},\ }\href {https://doi.org/10.1103/PhysRevLett.110.074801} {\bibfield
   {journal} {\bibinfo  {journal} {Phys. Rev. Lett.}\ }\textbf {\bibinfo
  {volume} {110}},\ \bibinfo {pages} {074801} (\bibinfo {year}
  {2013})}\BibitemShut {NoStop}%
\bibitem [{\citenamefont {Dowell}(2019)}]{dowell-2019a}%
  \BibitemOpen
  \bibfield  {author} {\bibinfo {author} {\bibfnamefont {D.~H.}\ \bibnamefont
  {Dowell}},\ }\bibfield  {title} {\bibinfo {title} {Topological cathodes:
  Controlling the space charge limit of electron emission using
  metamaterials},\ }\href {https://doi.org/10.1103/PhysRevAccelBeams.22.084201}
  {\bibfield  {journal} {\bibinfo  {journal} {Phys. Rev. Accel. Beams}\
  }\textbf {\bibinfo {volume} {22}},\ \bibinfo {pages} {084201} (\bibinfo
  {year} {2019})}\BibitemShut {NoStop}%
\bibitem [{\citenamefont {Musumeci}\ \emph {et~al.}(2008)\citenamefont
  {Musumeci}, \citenamefont {Moody}, \citenamefont {England}, \citenamefont
  {Rosenzweig},\ and\ \citenamefont {Tran}}]{musumeci-2008a}%
  \BibitemOpen
  \bibfield  {author} {\bibinfo {author} {\bibfnamefont {P.}~\bibnamefont
  {Musumeci}}, \bibinfo {author} {\bibfnamefont {J.~T.}\ \bibnamefont {Moody}},
  \bibinfo {author} {\bibfnamefont {R.~J.}\ \bibnamefont {England}}, \bibinfo
  {author} {\bibfnamefont {J.~B.}\ \bibnamefont {Rosenzweig}},\ and\ \bibinfo
  {author} {\bibfnamefont {T.}~\bibnamefont {Tran}},\ }\bibfield  {title}
  {\bibinfo {title} {Experimental generation and characterization of uniformly
  filled ellipsoidal electron-beam distributions},\ }\href
  {https://doi.org/10.1103/PhysRevLett.100.244801} {\bibfield  {journal}
  {\bibinfo  {journal} {Phys. Rev. Lett.}\ }\textbf {\bibinfo {volume} {100}},\
  \bibinfo {pages} {244801} (\bibinfo {year} {2008})}\BibitemShut {NoStop}%
\bibitem [{\citenamefont {Piot}\ \emph {et~al.}(2013)\citenamefont {Piot},
  \citenamefont {Sun}, \citenamefont {Maxwell}, \citenamefont {Ruan},
  \citenamefont {Secchi},\ and\ \citenamefont {Thangaraj}}]{piot-2013a}%
  \BibitemOpen
  \bibfield  {author} {\bibinfo {author} {\bibfnamefont {P.}~\bibnamefont
  {Piot}}, \bibinfo {author} {\bibfnamefont {Y.-E.}\ \bibnamefont {Sun}},
  \bibinfo {author} {\bibfnamefont {T.~J.}\ \bibnamefont {Maxwell}}, \bibinfo
  {author} {\bibfnamefont {J.}~\bibnamefont {Ruan}}, \bibinfo {author}
  {\bibfnamefont {E.}~\bibnamefont {Secchi}},\ and\ \bibinfo {author}
  {\bibfnamefont {J.~C.~T.}\ \bibnamefont {Thangaraj}},\ }\bibfield  {title}
  {\bibinfo {title} {Formation and acceleration of uniformly filled ellipsoidal
  electron bunches obtained via space-charge-driven expansion from a
  cesium-telluride photocathode},\ }\href
  {https://doi.org/10.1103/PhysRevSTAB.16.010102} {\bibfield  {journal}
  {\bibinfo  {journal} {Phys. Rev. ST Accel. Beams}\ }\textbf {\bibinfo
  {volume} {16}},\ \bibinfo {pages} {010102} (\bibinfo {year}
  {2013})}\BibitemShut {NoStop}%
\bibitem [{\citenamefont {Limborg-Deprey}\ and\ \citenamefont
  {Bolton}(2006)}]{limborg-2016a}%
  \BibitemOpen
  \bibfield  {author} {\bibinfo {author} {\bibfnamefont {C.}~\bibnamefont
  {Limborg-Deprey}}\ and\ \bibinfo {author} {\bibfnamefont {P.~R.}\
  \bibnamefont {Bolton}},\ }\bibfield  {title} {\bibinfo {title} {Optimum
  electron distributions for space charge dominated beams in photoinjectors},\
  }\href {https://doi.org/https://doi.org/10.1016/j.nima.2005.10.061}
  {\bibfield  {journal} {\bibinfo  {journal} {Nuclear Instruments and Methods
  in Physics Research Section A: Accelerators, Spectrometers, Detectors and
  Associated Equipment}\ }\textbf {\bibinfo {volume} {557}},\ \bibinfo {pages}
  {106} (\bibinfo {year} {2006})},\ \bibinfo {note} {energy Recovering Linacs
  2005}\BibitemShut {NoStop}%
\bibitem [{\citenamefont {Li}\ and\ \citenamefont {Lewellen}(2008)}]{li-2008a}%
  \BibitemOpen
  \bibfield  {author} {\bibinfo {author} {\bibfnamefont {Y.}~\bibnamefont
  {Li}}\ and\ \bibinfo {author} {\bibfnamefont {J.~W.}\ \bibnamefont
  {Lewellen}},\ }\bibfield  {title} {\bibinfo {title} {Generating a
  quasiellipsoidal electron beam by 3d laser-pulse shaping},\ }\href
  {https://doi.org/10.1103/PhysRevLett.100.074801} {\bibfield  {journal}
  {\bibinfo  {journal} {Phys. Rev. Lett.}\ }\textbf {\bibinfo {volume} {100}},\
  \bibinfo {pages} {074801} (\bibinfo {year} {2008})}\BibitemShut {NoStop}%
\bibitem [{\citenamefont {Kuzmin}\ \emph {et~al.}(2020)\citenamefont {Kuzmin},
  \citenamefont {Mironov}, \citenamefont {Gacheva}, \citenamefont {Potemkin},
  \citenamefont {Khazanov}, \citenamefont {Krasilnikov},\ and\ \citenamefont
  {Stephan}}]{kuzmin-2020a}%
  \BibitemOpen
  \bibfield  {author} {\bibinfo {author} {\bibfnamefont {I.~V.}\ \bibnamefont
  {Kuzmin}}, \bibinfo {author} {\bibfnamefont {S.~Y.}\ \bibnamefont {Mironov}},
  \bibinfo {author} {\bibfnamefont {E.~I.}\ \bibnamefont {Gacheva}}, \bibinfo
  {author} {\bibfnamefont {A.~K.}\ \bibnamefont {Potemkin}}, \bibinfo {author}
  {\bibfnamefont {E.~A.}\ \bibnamefont {Khazanov}}, \bibinfo {author}
  {\bibfnamefont {M.~A.}\ \bibnamefont {Krasilnikov}},\ and\ \bibinfo {author}
  {\bibfnamefont {F.}~\bibnamefont {Stephan}},\ }\bibfield  {title} {\bibinfo
  {title} {Shaping picosecond ellipsoidal laser pulses with periodic intensity
  modulation for electron photoinjectors},\ }\href
  {https://doi.org/10.1364/AO.383181} {\bibfield  {journal} {\bibinfo
  {journal} {Appl. Opt.}\ }\textbf {\bibinfo {volume} {59}},\ \bibinfo {pages}
  {2776} (\bibinfo {year} {2020})}\BibitemShut {NoStop}%
\bibitem [{\citenamefont {Bane}\ \emph {et~al.}(1985)\citenamefont {Bane},
  \citenamefont {Chen},\ and\ \citenamefont {Wilson}}]{bane-1985a}%
  \BibitemOpen
  \bibfield  {author} {\bibinfo {author} {\bibfnamefont {K.~L.}\ \bibnamefont
  {Bane}}, \bibinfo {author} {\bibfnamefont {P.}~\bibnamefont {Chen}},\ and\
  \bibinfo {author} {\bibfnamefont {P.~B.}\ \bibnamefont {Wilson}},\ }\bibfield
   {title} {\bibinfo {title} {{On collinear wake field acceleration}},\ }\href
  {https://doi.org/10.1109/TNS.1985.4334416} {\bibfield  {journal} {\bibinfo
  {journal} {IEEE Trans. Nucl. Sci.}\ }\textbf {\bibinfo {volume} {32}},\
  \bibinfo {pages} {3524} (\bibinfo {year} {1985})}\BibitemShut {NoStop}%
\bibitem [{\citenamefont {Tan}\ \emph {et~al.}(2021)\citenamefont {Tan},
  \citenamefont {Piot},\ and\ \citenamefont {Zholents}}]{tan-2021a}%
  \BibitemOpen
  \bibfield  {author} {\bibinfo {author} {\bibfnamefont {W.~H.}\ \bibnamefont
  {Tan}}, \bibinfo {author} {\bibfnamefont {P.}~\bibnamefont {Piot}},\ and\
  \bibinfo {author} {\bibfnamefont {A.}~\bibnamefont {Zholents}},\ }\bibfield
  {title} {\bibinfo {title} {Formation of temporally shaped electron bunches
  for beam-driven collinear wakefield accelerators},\ }\href
  {https://doi.org/10.1103/PhysRevAccelBeams.24.051303} {\bibfield  {journal}
  {\bibinfo  {journal} {Phys. Rev. Accel. Beams}\ }\textbf {\bibinfo {volume}
  {24}},\ \bibinfo {pages} {051303} (\bibinfo {year} {2021})}\BibitemShut
  {NoStop}%
\bibitem [{\citenamefont {Xu}\ \emph {et~al.}(2019)\citenamefont {Xu},
  \citenamefont {Jing}, \citenamefont {Kanareykin}, \citenamefont {Piot},\ and\
  \citenamefont {Power}}]{xu-2019a}%
  \BibitemOpen
  \bibfield  {author} {\bibinfo {author} {\bibfnamefont {T.}~\bibnamefont
  {Xu}}, \bibinfo {author} {\bibfnamefont {C.-J.}\ \bibnamefont {Jing}},
  \bibinfo {author} {\bibfnamefont {A.}~\bibnamefont {Kanareykin}}, \bibinfo
  {author} {\bibfnamefont {P.}~\bibnamefont {Piot}},\ and\ \bibinfo {author}
  {\bibfnamefont {J.}~\bibnamefont {Power}},\ }\bibfield  {title} {\bibinfo
  {title} {{S}patio{-T}emporal {S}haping of the {P}hotocathode {L}aser {P}ulse
  for {L}ow{-E}mittance {S}haped {E}lectron {B}unches},\ }in\ \href
  {https://doi.org/doi:10.18429/JACoW-IPAC2019-TUPTS104} {\emph {\bibinfo
  {booktitle} {Proc. 10th International Particle Accelerator Conference
  (IPAC'19), Melbourne, Australia, 19-24 May 2019}}},\ \bibinfo {series and
  number} {\bibinfo {series} {International Particle Accelerator Conference}\
  No.~\bibinfo {number} {10}}\ (\bibinfo  {publisher} {JACoW Publishing},\
  \bibinfo {address} {Geneva, Switzerland},\ \bibinfo {year} {2019})\ pp.\
  \bibinfo {pages} {2163--2166}\BibitemShut {NoStop}%
\bibitem [{\citenamefont {Petrushina}\ \emph {et~al.}(2020)\citenamefont
  {Petrushina}, \citenamefont {Litvinenko}, \citenamefont {Jing}, \citenamefont
  {Ma}, \citenamefont {Pinayev}, \citenamefont {Shih}, \citenamefont {Wang},
  \citenamefont {Wu}, \citenamefont {Altinbas}, \citenamefont {Brutus},
  \citenamefont {Belomestnykh}, \citenamefont {Di~Lieto}, \citenamefont
  {Inacker}, \citenamefont {Jamilkowski}, \citenamefont {Mahler}, \citenamefont
  {Mapes}, \citenamefont {Miller}, \citenamefont {Narayan}, \citenamefont
  {Paniccia}, \citenamefont {Roser}, \citenamefont {Severino}, \citenamefont
  {Skaritka}, \citenamefont {Smart}, \citenamefont {Smith}, \citenamefont
  {Soria}, \citenamefont {Than}, \citenamefont {Tuozzolo}, \citenamefont
  {Wang}, \citenamefont {Xiao}, \citenamefont {Xin}, \citenamefont {Ben-Zvi},
  \citenamefont {Boulware}, \citenamefont {Grimm}, \citenamefont {Mihara},
  \citenamefont {Kayran},\ and\ \citenamefont {Rao}}]{PhysRevLett.124.244801}%
  \BibitemOpen
  \bibfield  {author} {\bibinfo {author} {\bibfnamefont {I.}~\bibnamefont
  {Petrushina}}, \bibinfo {author} {\bibfnamefont {V.~N.}\ \bibnamefont
  {Litvinenko}}, \bibinfo {author} {\bibfnamefont {Y.}~\bibnamefont {Jing}},
  \bibinfo {author} {\bibfnamefont {J.}~\bibnamefont {Ma}}, \bibinfo {author}
  {\bibfnamefont {I.}~\bibnamefont {Pinayev}}, \bibinfo {author} {\bibfnamefont
  {K.}~\bibnamefont {Shih}}, \bibinfo {author} {\bibfnamefont {G.}~\bibnamefont
  {Wang}}, \bibinfo {author} {\bibfnamefont {Y.~H.}\ \bibnamefont {Wu}},
  \bibinfo {author} {\bibfnamefont {Z.}~\bibnamefont {Altinbas}}, \bibinfo
  {author} {\bibfnamefont {J.~C.}\ \bibnamefont {Brutus}}, \bibinfo {author}
  {\bibfnamefont {S.}~\bibnamefont {Belomestnykh}}, \bibinfo {author}
  {\bibfnamefont {A.}~\bibnamefont {Di~Lieto}}, \bibinfo {author}
  {\bibfnamefont {P.}~\bibnamefont {Inacker}}, \bibinfo {author} {\bibfnamefont
  {J.}~\bibnamefont {Jamilkowski}}, \bibinfo {author} {\bibfnamefont
  {G.}~\bibnamefont {Mahler}}, \bibinfo {author} {\bibfnamefont
  {M.}~\bibnamefont {Mapes}}, \bibinfo {author} {\bibfnamefont
  {T.}~\bibnamefont {Miller}}, \bibinfo {author} {\bibfnamefont
  {G.}~\bibnamefont {Narayan}}, \bibinfo {author} {\bibfnamefont
  {M.}~\bibnamefont {Paniccia}}, \bibinfo {author} {\bibfnamefont
  {T.}~\bibnamefont {Roser}}, \bibinfo {author} {\bibfnamefont
  {F.}~\bibnamefont {Severino}}, \bibinfo {author} {\bibfnamefont
  {J.}~\bibnamefont {Skaritka}}, \bibinfo {author} {\bibfnamefont
  {L.}~\bibnamefont {Smart}}, \bibinfo {author} {\bibfnamefont
  {K.}~\bibnamefont {Smith}}, \bibinfo {author} {\bibfnamefont
  {V.}~\bibnamefont {Soria}}, \bibinfo {author} {\bibfnamefont
  {Y.}~\bibnamefont {Than}}, \bibinfo {author} {\bibfnamefont {J.}~\bibnamefont
  {Tuozzolo}}, \bibinfo {author} {\bibfnamefont {E.}~\bibnamefont {Wang}},
  \bibinfo {author} {\bibfnamefont {B.}~\bibnamefont {Xiao}}, \bibinfo {author}
  {\bibfnamefont {T.}~\bibnamefont {Xin}}, \bibinfo {author} {\bibfnamefont
  {I.}~\bibnamefont {Ben-Zvi}}, \bibinfo {author} {\bibfnamefont
  {C.}~\bibnamefont {Boulware}}, \bibinfo {author} {\bibfnamefont
  {T.}~\bibnamefont {Grimm}}, \bibinfo {author} {\bibfnamefont
  {K.}~\bibnamefont {Mihara}}, \bibinfo {author} {\bibfnamefont
  {D.}~\bibnamefont {Kayran}},\ and\ \bibinfo {author} {\bibfnamefont
  {T.}~\bibnamefont {Rao}},\ }\bibfield  {title} {\bibinfo {title}
  {High-brightness continuous-wave electron beams from superconducting
  radio-frequency photoemission gun},\ }\href
  {https://doi.org/10.1103/PhysRevLett.124.244801} {\bibfield  {journal}
  {\bibinfo  {journal} {Phys. Rev. Lett.}\ }\textbf {\bibinfo {volume} {124}},\
  \bibinfo {pages} {244801} (\bibinfo {year} {2020})}\BibitemShut {NoStop}%
\bibitem [{\citenamefont {Musumeci}\ \emph {et~al.}(2018)\citenamefont
  {Musumeci}, \citenamefont {Navarro}, \citenamefont {Rosenzweig},
  \citenamefont {Cultrera}, \citenamefont {Bazarov}, \citenamefont {Maxson},
  \citenamefont {Karkare},\ and\ \citenamefont
  {Padmore}}]{musumeci2018advances}%
  \BibitemOpen
  \bibfield  {author} {\bibinfo {author} {\bibfnamefont {P.}~\bibnamefont
  {Musumeci}}, \bibinfo {author} {\bibfnamefont {J.~G.}\ \bibnamefont
  {Navarro}}, \bibinfo {author} {\bibfnamefont {J.}~\bibnamefont {Rosenzweig}},
  \bibinfo {author} {\bibfnamefont {L.}~\bibnamefont {Cultrera}}, \bibinfo
  {author} {\bibfnamefont {I.}~\bibnamefont {Bazarov}}, \bibinfo {author}
  {\bibfnamefont {J.}~\bibnamefont {Maxson}}, \bibinfo {author} {\bibfnamefont
  {S.}~\bibnamefont {Karkare}},\ and\ \bibinfo {author} {\bibfnamefont
  {H.}~\bibnamefont {Padmore}},\ }\bibfield  {title} {\bibinfo {title}
  {Advances in bright electron sources},\ }\href@noop {} {\bibfield  {journal}
  {\bibinfo  {journal} {Nuclear Instruments and Methods in Physics Research
  Section A: Accelerators, Spectrometers, Detectors and Associated Equipment}\
  }\textbf {\bibinfo {volume} {907}},\ \bibinfo {pages} {209} (\bibinfo {year}
  {2018})}\BibitemShut {NoStop}%
\bibitem [{\citenamefont {Maruyama}\ \emph {et~al.}(1992)\citenamefont
  {Maruyama}, \citenamefont {Garwin}, \citenamefont {Prepost},\ and\
  \citenamefont {Zapalac}}]{Maruyama1992}%
  \BibitemOpen
  \bibfield  {author} {\bibinfo {author} {\bibfnamefont {T.}~\bibnamefont
  {Maruyama}}, \bibinfo {author} {\bibfnamefont {E.~L.}\ \bibnamefont
  {Garwin}}, \bibinfo {author} {\bibfnamefont {R.}~\bibnamefont {Prepost}},\
  and\ \bibinfo {author} {\bibfnamefont {G.~H.}\ \bibnamefont {Zapalac}},\
  }\bibfield  {title} {\bibinfo {title} {Electron-spin polarization in
  photoemission from strained gaas grown on
  ${\mathrm{gaas}}_{1\mathrm{\ensuremath{-}}\mathit{x}}$${\mathrm{p}}_{\mathit{x}}$},\
  }\href {https://doi.org/10.1103/PhysRevB.46.4261} {\bibfield  {journal}
  {\bibinfo  {journal} {Phys. Rev. B}\ }\textbf {\bibinfo {volume} {46}},\
  \bibinfo {pages} {4261} (\bibinfo {year} {1992})}\BibitemShut {NoStop}%
\bibitem [{\citenamefont {Liu}\ \emph {et~al.}(2016)\citenamefont {Liu},
  \citenamefont {Chen}, \citenamefont {Lu}, \citenamefont {Moy}, \citenamefont
  {Poelker}, \citenamefont {Stutzman},\ and\ \citenamefont {Zhang}}]{Liu2016}%
  \BibitemOpen
  \bibfield  {author} {\bibinfo {author} {\bibfnamefont {W.}~\bibnamefont
  {Liu}}, \bibinfo {author} {\bibfnamefont {Y.}~\bibnamefont {Chen}}, \bibinfo
  {author} {\bibfnamefont {W.}~\bibnamefont {Lu}}, \bibinfo {author}
  {\bibfnamefont {A.}~\bibnamefont {Moy}}, \bibinfo {author} {\bibfnamefont
  {M.}~\bibnamefont {Poelker}}, \bibinfo {author} {\bibfnamefont
  {M.}~\bibnamefont {Stutzman}},\ and\ \bibinfo {author} {\bibfnamefont
  {S.}~\bibnamefont {Zhang}},\ }\bibfield  {title} {\bibinfo {title}
  {Record-level quantum efficiency from a high polarization strained gaas/gaasp
  superlattice photocathode with distributed bragg reflector},\ }\href
  {https://doi.org/10.1063/1.4972180} {\bibfield  {journal} {\bibinfo
  {journal} {Applied Physics Letters}\ }\textbf {\bibinfo {volume} {109}},\
  \bibinfo {pages} {252104} (\bibinfo {year} {2016})},\ \Eprint
  {https://arxiv.org/abs/https://doi.org/10.1063/1.4972180}
  {https://doi.org/10.1063/1.4972180} \BibitemShut {NoStop}%
\bibitem [{\citenamefont {Dunham}\ \emph {et~al.}(2013)\citenamefont {Dunham},
  \citenamefont {Barley}, \citenamefont {Bartnik}, \citenamefont {Bazarov},
  \citenamefont {Cultrera}, \citenamefont {Dobbins}, \citenamefont
  {Hoffstaetter}, \citenamefont {Johnson}, \citenamefont {Kaplan},
  \citenamefont {Karkare} \emph {et~al.}}]{dunham2013record}%
  \BibitemOpen
  \bibfield  {author} {\bibinfo {author} {\bibfnamefont {B.}~\bibnamefont
  {Dunham}}, \bibinfo {author} {\bibfnamefont {J.}~\bibnamefont {Barley}},
  \bibinfo {author} {\bibfnamefont {A.}~\bibnamefont {Bartnik}}, \bibinfo
  {author} {\bibfnamefont {I.}~\bibnamefont {Bazarov}}, \bibinfo {author}
  {\bibfnamefont {L.}~\bibnamefont {Cultrera}}, \bibinfo {author}
  {\bibfnamefont {J.}~\bibnamefont {Dobbins}}, \bibinfo {author} {\bibfnamefont
  {G.}~\bibnamefont {Hoffstaetter}}, \bibinfo {author} {\bibfnamefont
  {B.}~\bibnamefont {Johnson}}, \bibinfo {author} {\bibfnamefont
  {R.}~\bibnamefont {Kaplan}}, \bibinfo {author} {\bibfnamefont
  {S.}~\bibnamefont {Karkare}}, \emph {et~al.},\ }\bibfield  {title} {\bibinfo
  {title} {Record high-average current from a high-brightness photoinjector},\
  }\href@noop {} {\bibfield  {journal} {\bibinfo  {journal} {Applied Physics
  Letters}\ }\textbf {\bibinfo {volume} {102}},\ \bibinfo {pages} {034105}
  (\bibinfo {year} {2013})}\BibitemShut {NoStop}%
\bibitem [{\citenamefont {Filippetto}\ \emph
  {et~al.}(2015{\natexlab{a}})\citenamefont {Filippetto}, \citenamefont
  {Qian},\ and\ \citenamefont {Sannibale}}]{Filippetto2015}%
  \BibitemOpen
  \bibfield  {author} {\bibinfo {author} {\bibfnamefont {D.}~\bibnamefont
  {Filippetto}}, \bibinfo {author} {\bibfnamefont {H.}~\bibnamefont {Qian}},\
  and\ \bibinfo {author} {\bibfnamefont {F.}~\bibnamefont {Sannibale}},\
  }\bibfield  {title} {\bibinfo {title} {Cesium telluride cathodes for the next
  generation of high-average current high-brightness photoinjectors},\ }\href
  {https://doi.org/10.1063/1.4927700} {\bibfield  {journal} {\bibinfo
  {journal} {Applied Physics Letters}\ }\textbf {\bibinfo {volume} {107}},\
  \bibinfo {pages} {042104} (\bibinfo {year} {2015}{\natexlab{a}})},\ \Eprint
  {https://arxiv.org/abs/https://doi.org/10.1063/1.4927700}
  {https://doi.org/10.1063/1.4927700} \BibitemShut {NoStop}%
\bibitem [{\citenamefont {Antoniuk}\ \emph {et~al.}(2021)\citenamefont
  {Antoniuk}, \citenamefont {Schindler}, \citenamefont {Schroeder},
  \citenamefont {Dunham}, \citenamefont {Pianetta}, \citenamefont {Vecchione},\
  and\ \citenamefont {Reed}}]{Antoniuk2021}%
  \BibitemOpen
  \bibfield  {author} {\bibinfo {author} {\bibfnamefont {E.~R.}\ \bibnamefont
  {Antoniuk}}, \bibinfo {author} {\bibfnamefont {P.}~\bibnamefont {Schindler}},
  \bibinfo {author} {\bibfnamefont {W.~A.}\ \bibnamefont {Schroeder}}, \bibinfo
  {author} {\bibfnamefont {B.}~\bibnamefont {Dunham}}, \bibinfo {author}
  {\bibfnamefont {P.}~\bibnamefont {Pianetta}}, \bibinfo {author}
  {\bibfnamefont {T.}~\bibnamefont {Vecchione}},\ and\ \bibinfo {author}
  {\bibfnamefont {E.~J.}\ \bibnamefont {Reed}},\ }\bibfield  {title} {\bibinfo
  {title} {Novel ultrabright and air-stable photocathodes discovered from
  machine learning and density functional theory driven screening},\ }\href
  {https://doi.org/https://doi.org/10.1002/adma.202104081} {\bibfield
  {journal} {\bibinfo  {journal} {Advanced Materials}\ }\textbf {\bibinfo
  {volume} {33}},\ \bibinfo {pages} {2104081} (\bibinfo {year} {2021})},\
  \Eprint
  {https://arxiv.org/abs/https://onlinelibrary.wiley.com/doi/pdf/10.1002/adma.202104081}
  {https://onlinelibrary.wiley.com/doi/pdf/10.1002/adma.202104081} \BibitemShut
  {NoStop}%
\bibitem [{\citenamefont {Chubenko}\ \emph {et~al.}(2021)\citenamefont
  {Chubenko}, \citenamefont {Karkare}, \citenamefont {Dimitrov}, \citenamefont
  {Bae}, \citenamefont {Cultrera}, \citenamefont {Bazarov},\ and\ \citenamefont
  {Afanasev}}]{chubenko2021monte}%
  \BibitemOpen
  \bibfield  {author} {\bibinfo {author} {\bibfnamefont {O.}~\bibnamefont
  {Chubenko}}, \bibinfo {author} {\bibfnamefont {S.}~\bibnamefont {Karkare}},
  \bibinfo {author} {\bibfnamefont {D.~A.}\ \bibnamefont {Dimitrov}}, \bibinfo
  {author} {\bibfnamefont {J.~K.}\ \bibnamefont {Bae}}, \bibinfo {author}
  {\bibfnamefont {L.}~\bibnamefont {Cultrera}}, \bibinfo {author}
  {\bibfnamefont {I.}~\bibnamefont {Bazarov}},\ and\ \bibinfo {author}
  {\bibfnamefont {A.}~\bibnamefont {Afanasev}},\ }\bibfield  {title} {\bibinfo
  {title} {Monte carlo modeling of spin-polarized photoemission from p-doped
  bulk gaas},\ }\href@noop {} {\bibfield  {journal} {\bibinfo  {journal}
  {Journal of Applied Physics}\ }\textbf {\bibinfo {volume} {130}},\ \bibinfo
  {pages} {063101} (\bibinfo {year} {2021})}\BibitemShut {NoStop}%
\bibitem [{\citenamefont {Bai}\ \emph {et~al.}(2021)\citenamefont {Bai},
  \citenamefont {Barklow}, \citenamefont {Bartoldus}, \citenamefont
  {Breidenbach}, \citenamefont {Grenier}, \citenamefont {Huang}, \citenamefont
  {Kagan}, \citenamefont {Lewellen}, \citenamefont {Li}, \citenamefont
  {Markiewicz}, \citenamefont {Nanni}, \citenamefont {Nasr}, \citenamefont
  {Ng}, \citenamefont {Oriunno}, \citenamefont {Peskin}, \citenamefont {Rizzo},
  \citenamefont {Rosenzweig}, \citenamefont {Schwartzman}, \citenamefont
  {Shiltsev}, \citenamefont {Simakov}, \citenamefont {Spataro}, \citenamefont
  {Su}, \citenamefont {Tantawi}, \citenamefont {Vernieri}, \citenamefont
  {White},\ and\ \citenamefont {Young}}]{breidenbach_ccc_nodate}%
  \BibitemOpen
  \bibfield  {author} {\bibinfo {author} {\bibfnamefont {M.}~\bibnamefont
  {Bai}}, \bibinfo {author} {\bibfnamefont {T.}~\bibnamefont {Barklow}},
  \bibinfo {author} {\bibfnamefont {R.}~\bibnamefont {Bartoldus}}, \bibinfo
  {author} {\bibfnamefont {M.}~\bibnamefont {Breidenbach}}, \bibinfo {author}
  {\bibfnamefont {P.}~\bibnamefont {Grenier}}, \bibinfo {author} {\bibfnamefont
  {Z.}~\bibnamefont {Huang}}, \bibinfo {author} {\bibfnamefont
  {M.}~\bibnamefont {Kagan}}, \bibinfo {author} {\bibfnamefont
  {J.}~\bibnamefont {Lewellen}}, \bibinfo {author} {\bibfnamefont
  {Z.}~\bibnamefont {Li}}, \bibinfo {author} {\bibfnamefont {T.~W.}\
  \bibnamefont {Markiewicz}}, \bibinfo {author} {\bibfnamefont {E.~A.}\
  \bibnamefont {Nanni}}, \bibinfo {author} {\bibfnamefont {M.}~\bibnamefont
  {Nasr}}, \bibinfo {author} {\bibfnamefont {C.-K.}\ \bibnamefont {Ng}},
  \bibinfo {author} {\bibfnamefont {M.}~\bibnamefont {Oriunno}}, \bibinfo
  {author} {\bibfnamefont {M.~E.}\ \bibnamefont {Peskin}}, \bibinfo {author}
  {\bibfnamefont {T.~G.}\ \bibnamefont {Rizzo}}, \bibinfo {author}
  {\bibfnamefont {J.}~\bibnamefont {Rosenzweig}}, \bibinfo {author}
  {\bibfnamefont {A.~G.}\ \bibnamefont {Schwartzman}}, \bibinfo {author}
  {\bibfnamefont {V.}~\bibnamefont {Shiltsev}}, \bibinfo {author}
  {\bibfnamefont {E.}~\bibnamefont {Simakov}}, \bibinfo {author} {\bibfnamefont
  {B.}~\bibnamefont {Spataro}}, \bibinfo {author} {\bibfnamefont
  {D.}~\bibnamefont {Su}}, \bibinfo {author} {\bibfnamefont {S.}~\bibnamefont
  {Tantawi}}, \bibinfo {author} {\bibfnamefont {C.}~\bibnamefont {Vernieri}},
  \bibinfo {author} {\bibfnamefont {G.}~\bibnamefont {White}},\ and\ \bibinfo
  {author} {\bibfnamefont {C.~C.}\ \bibnamefont {Young}},\ }\href
  {https://doi.org/10.48550/ARXIV.2110.15800} {\bibinfo {title} {C$^3$: A
  "cool" route to the higgs boson and beyond}} (\bibinfo {year}
  {2021})\BibitemShut {NoStop}%
\bibitem [{\citenamefont {Charles}\ \emph {et~al.}(2018)\citenamefont {Charles}
  \emph {et~al.}}]{CLIC}%
  \BibitemOpen
  \bibfield  {author} {\bibinfo {author} {\bibfnamefont {T.~K.}\ \bibnamefont
  {Charles}} \emph {et~al.},\ }\href {https://doi.org/10.23731/CYRM-2018-002}
  {\bibinfo {title} {The compact linear collider (clic) - 2018 summary report}}
  (\bibinfo {year} {2018}),\ \Eprint {https://arxiv.org/abs/1812.06018}
  {arXiv:1812.06018 [physics.acc-ph]} \BibitemShut {NoStop}%
\bibitem [{\citenamefont {Nghiem}\ \emph {et~al.}(2020)\citenamefont {Nghiem},
  \citenamefont {Assmann}, \citenamefont {Beck}, \citenamefont {Chanc\'e},
  \citenamefont {Chiadroni}, \citenamefont {Cros}, \citenamefont {Ferrario},
  \citenamefont {Ferran~Pousa}, \citenamefont {Giribono}, \citenamefont
  {Gizzi}, \citenamefont {Hidding}, \citenamefont {Lee}, \citenamefont {Li},
  \citenamefont {Marocchino}, \citenamefont {Martinez de~la Ossa},
  \citenamefont {Massimo}, \citenamefont {Maynard}, \citenamefont {Mosnier},
  \citenamefont {Romeo}, \citenamefont {Rossi}, \citenamefont {Silva},
  \citenamefont {Svystun}, \citenamefont {Tomassini}, \citenamefont
  {Vaccarezza}, \citenamefont {Vieira},\ and\ \citenamefont
  {Zhu}}]{nghiem_toward_2020}%
  \BibitemOpen
  \bibfield  {author} {\bibinfo {author} {\bibfnamefont {P.~A.~P.}\
  \bibnamefont {Nghiem}}, \bibinfo {author} {\bibfnamefont {R.}~\bibnamefont
  {Assmann}}, \bibinfo {author} {\bibfnamefont {A.}~\bibnamefont {Beck}},
  \bibinfo {author} {\bibfnamefont {A.}~\bibnamefont {Chanc\'e}}, \bibinfo
  {author} {\bibfnamefont {E.}~\bibnamefont {Chiadroni}}, \bibinfo {author}
  {\bibfnamefont {B.}~\bibnamefont {Cros}}, \bibinfo {author} {\bibfnamefont
  {M.}~\bibnamefont {Ferrario}}, \bibinfo {author} {\bibfnamefont
  {A.}~\bibnamefont {Ferran~Pousa}}, \bibinfo {author} {\bibfnamefont
  {A.}~\bibnamefont {Giribono}}, \bibinfo {author} {\bibfnamefont {L.~A.}\
  \bibnamefont {Gizzi}}, \bibinfo {author} {\bibfnamefont {B.}~\bibnamefont
  {Hidding}}, \bibinfo {author} {\bibfnamefont {P.}~\bibnamefont {Lee}},
  \bibinfo {author} {\bibfnamefont {X.}~\bibnamefont {Li}}, \bibinfo {author}
  {\bibfnamefont {A.}~\bibnamefont {Marocchino}}, \bibinfo {author}
  {\bibfnamefont {A.}~\bibnamefont {Martinez de~la Ossa}}, \bibinfo {author}
  {\bibfnamefont {F.}~\bibnamefont {Massimo}}, \bibinfo {author} {\bibfnamefont
  {G.}~\bibnamefont {Maynard}}, \bibinfo {author} {\bibfnamefont
  {A.}~\bibnamefont {Mosnier}}, \bibinfo {author} {\bibfnamefont
  {S.}~\bibnamefont {Romeo}}, \bibinfo {author} {\bibfnamefont {A.~R.}\
  \bibnamefont {Rossi}}, \bibinfo {author} {\bibfnamefont {T.}~\bibnamefont
  {Silva}}, \bibinfo {author} {\bibfnamefont {E.}~\bibnamefont {Svystun}},
  \bibinfo {author} {\bibfnamefont {P.}~\bibnamefont {Tomassini}}, \bibinfo
  {author} {\bibfnamefont {C.}~\bibnamefont {Vaccarezza}}, \bibinfo {author}
  {\bibfnamefont {J.}~\bibnamefont {Vieira}},\ and\ \bibinfo {author}
  {\bibfnamefont {J.}~\bibnamefont {Zhu}},\ }\bibfield  {title} {\bibinfo
  {title} {Toward a plasma-based accelerator at high beam energy with high beam
  charge and high beam quality},\ }\href
  {https://doi.org/10.1103/PhysRevAccelBeams.23.031301} {\bibfield  {journal}
  {\bibinfo  {journal} {Phys. Rev. Accel. Beams}\ }\textbf {\bibinfo {volume}
  {23}},\ \bibinfo {pages} {031301} (\bibinfo {year} {2020})}\BibitemShut
  {NoStop}%
\bibitem [{\citenamefont {Willeke}\ and\ \citenamefont {Beebe-Wang}()}]{EIC}%
  \BibitemOpen
  \bibfield  {author} {\bibinfo {author} {\bibfnamefont {F.}~\bibnamefont
  {Willeke}}\ and\ \bibinfo {author} {\bibfnamefont {J.}~\bibnamefont
  {Beebe-Wang}},\ }\href
  {https://www.osti.gov/biblio/1765663-electron-ion-collider-conceptual-design-report}
  {\bibinfo {title} {Electron ion collider conceptual design report
  2021}}\BibitemShut {NoStop}%
\bibitem [{\citenamefont {Piot}\ \emph
  {et~al.}(2006{\natexlab{a}})\citenamefont {Piot}, \citenamefont {Sun},\ and\
  \citenamefont {Kim}}]{piot_photoinjector_2006}%
  \BibitemOpen
  \bibfield  {author} {\bibinfo {author} {\bibfnamefont {P.}~\bibnamefont
  {Piot}}, \bibinfo {author} {\bibfnamefont {Y.-E.}\ \bibnamefont {Sun}},\ and\
  \bibinfo {author} {\bibfnamefont {K.-J.}\ \bibnamefont {Kim}},\ }\bibfield
  {title} {\bibinfo {title} {Photoinjector generation of a flat electron beam
  with transverse emittance ratio of 100},\ }\bibfield  {journal} {\bibinfo
  {journal} {Physical Review Special Topics - Accelerators and Beams}\ }\textbf
  {\bibinfo {volume} {9}},\ \href
  {https://doi.org/10.1103/PhysRevSTAB.9.031001} {10.1103/PhysRevSTAB.9.031001}
  (\bibinfo {year} {2006}{\natexlab{a}})\BibitemShut {NoStop}%
\bibitem [{\citenamefont {Grames}\ \emph {et~al.}(2018)\citenamefont {Grames},
  \citenamefont {Adderley}, \citenamefont {Hansknecht}, \citenamefont
  {Poelker}, \citenamefont {Moser}, \citenamefont {Stutzman},\ and\
  \citenamefont {Zhang}}]{grames2018milliampere}%
  \BibitemOpen
  \bibfield  {author} {\bibinfo {author} {\bibfnamefont {J.~M.}\ \bibnamefont
  {Grames}}, \bibinfo {author} {\bibfnamefont {P.~A.}\ \bibnamefont
  {Adderley}}, \bibinfo {author} {\bibfnamefont {J.~C.}\ \bibnamefont
  {Hansknecht}}, \bibinfo {author} {\bibfnamefont {M.}~\bibnamefont {Poelker}},
  \bibinfo {author} {\bibfnamefont {D.~G.}\ \bibnamefont {Moser}}, \bibinfo
  {author} {\bibfnamefont {M.~L.}\ \bibnamefont {Stutzman}},\ and\ \bibinfo
  {author} {\bibfnamefont {S.}~\bibnamefont {Zhang}},\ }\href@noop {} {\emph
  {\bibinfo {title} {Milliampere beam studies using high polarization
  photocathodes at the CEBAF Photoinjector}}},\ \bibinfo {type} {Tech. Rep.}\
  (\bibinfo  {institution} {Thomas Jefferson National Accelerator Facility
  (TJNAF), Newport News, VA~…},\ \bibinfo {year} {2018})\BibitemShut
  {NoStop}%
\bibitem [{\citenamefont {Friederich}\ \emph {et~al.}(2021)\citenamefont
  {Friederich}, \citenamefont {Aulenbacher},\ and\ \citenamefont
  {Matejcek}}]{friederichstatus}%
  \BibitemOpen
  \bibfield  {author} {\bibinfo {author} {\bibfnamefont {S.}~\bibnamefont
  {Friederich}}, \bibinfo {author} {\bibfnamefont {K.}~\bibnamefont
  {Aulenbacher}},\ and\ \bibinfo {author} {\bibfnamefont {C.}~\bibnamefont
  {Matejcek}},\ }\bibfield  {title} {\bibinfo {title} {{Status of the Polarized
  Source and Beam Preparation System at MESA}},\ }in\ \href
  {https://jacow.org/ipac2021/papers/wepab063.pdf} {\emph {\bibinfo {booktitle}
  {Proc. IPAC'21}}},\ \bibinfo {series and number} {\bibinfo {series}
  {International Particle Accelerator Conference}\ No.~\bibinfo {number} {12}}\
  (\bibinfo  {publisher} {JACoW Publishing, Geneva, Switzerland},\ \bibinfo
  {year} {2021})\ pp.\ \bibinfo {pages} {2736--2739}\BibitemShut {NoStop}%
\bibitem [{\citenamefont {Wang}\ \emph {et~al.}(2022)\citenamefont {Wang},
  \citenamefont {Rahman}, \citenamefont {Skaritka}, \citenamefont {Liu},
  \citenamefont {Biswas}, \citenamefont {Degen}, \citenamefont {Inacker},
  \citenamefont {Lambiase},\ and\ \citenamefont {Paniccia}}]{wang2022high}%
  \BibitemOpen
  \bibfield  {author} {\bibinfo {author} {\bibfnamefont {E.}~\bibnamefont
  {Wang}}, \bibinfo {author} {\bibfnamefont {O.}~\bibnamefont {Rahman}},
  \bibinfo {author} {\bibfnamefont {J.}~\bibnamefont {Skaritka}}, \bibinfo
  {author} {\bibfnamefont {W.}~\bibnamefont {Liu}}, \bibinfo {author}
  {\bibfnamefont {J.}~\bibnamefont {Biswas}}, \bibinfo {author} {\bibfnamefont
  {C.}~\bibnamefont {Degen}}, \bibinfo {author} {\bibfnamefont
  {P.}~\bibnamefont {Inacker}}, \bibinfo {author} {\bibfnamefont
  {R.}~\bibnamefont {Lambiase}},\ and\ \bibinfo {author} {\bibfnamefont
  {M.}~\bibnamefont {Paniccia}},\ }\bibfield  {title} {\bibinfo {title} {High
  voltage dc gun for high intensity polarized electron source},\ }\href@noop {}
  {\bibfield  {journal} {\bibinfo  {journal} {Physical Review Accelerators and
  Beams}\ }\textbf {\bibinfo {volume} {25}},\ \bibinfo {pages} {033401}
  (\bibinfo {year} {2022})}\BibitemShut {NoStop}%
\bibitem [{\citenamefont {Suleiman}\ \emph {et~al.}(2018)\citenamefont
  {Suleiman}, \citenamefont {Adderley}, \citenamefont {Grames}, \citenamefont
  {Hansknecht}, \citenamefont {Poelker},\ and\ \citenamefont
  {Stutzman}}]{suleiman2018high}%
  \BibitemOpen
  \bibfield  {author} {\bibinfo {author} {\bibfnamefont {R.}~\bibnamefont
  {Suleiman}}, \bibinfo {author} {\bibfnamefont {P.}~\bibnamefont {Adderley}},
  \bibinfo {author} {\bibfnamefont {J.}~\bibnamefont {Grames}}, \bibinfo
  {author} {\bibfnamefont {J.}~\bibnamefont {Hansknecht}}, \bibinfo {author}
  {\bibfnamefont {M.}~\bibnamefont {Poelker}},\ and\ \bibinfo {author}
  {\bibfnamefont {M.}~\bibnamefont {Stutzman}},\ }\bibfield  {title} {\bibinfo
  {title} {High current polarized electron source},\ }in\ \href@noop {} {\emph
  {\bibinfo {booktitle} {AIP Conference Proceedings}}},\ Vol.\ \bibinfo
  {volume} {1970}\ (\bibinfo {organization} {AIP Publishing LLC},\ \bibinfo
  {year} {2018})\ p.\ \bibinfo {pages} {050007}\BibitemShut {NoStop}%
\bibitem [{\citenamefont {Yoskowitz}\ \emph {et~al.}(2021)\citenamefont
  {Yoskowitz}, \citenamefont {Krafft}, \citenamefont {Palacios~Serrano},
  \citenamefont {Wijethunga}, \citenamefont {Grames}, \citenamefont
  {Hansknecht}, \citenamefont {Hernandez-Garcia}, \citenamefont {Poelker},
  \citenamefont {Stutzman}, \citenamefont {Suleiman} \emph
  {et~al.}}]{yoskowitz2021improving}%
  \BibitemOpen
  \bibfield  {author} {\bibinfo {author} {\bibfnamefont {J.}~\bibnamefont
  {Yoskowitz}}, \bibinfo {author} {\bibfnamefont {G.}~\bibnamefont {Krafft}},
  \bibinfo {author} {\bibfnamefont {G.}~\bibnamefont {Palacios~Serrano}},
  \bibinfo {author} {\bibfnamefont {S.}~\bibnamefont {Wijethunga}}, \bibinfo
  {author} {\bibfnamefont {J.}~\bibnamefont {Grames}}, \bibinfo {author}
  {\bibfnamefont {J.}~\bibnamefont {Hansknecht}}, \bibinfo {author}
  {\bibfnamefont {C.}~\bibnamefont {Hernandez-Garcia}}, \bibinfo {author}
  {\bibfnamefont {M.}~\bibnamefont {Poelker}}, \bibinfo {author} {\bibfnamefont
  {M.}~\bibnamefont {Stutzman}}, \bibinfo {author} {\bibfnamefont
  {R.}~\bibnamefont {Suleiman}}, \emph {et~al.},\ }\href@noop {} {\emph
  {\bibinfo {title} {Improving the Operational Lifetime of the CEBAF Photo-Gun
  by Anode Biasing}}},\ \bibinfo {type} {Tech. Rep.}\ (\bibinfo  {institution}
  {Thomas Jefferson National Accelerator Facility (TJNAF), Newport News,
  VA~…},\ \bibinfo {year} {2021})\BibitemShut {NoStop}%
\bibitem [{\citenamefont {Kayran}(2018)}]{kayran2018lerec}%
  \BibitemOpen
  \bibfield  {author} {\bibinfo {author} {\bibfnamefont {D.}~\bibnamefont
  {Kayran}},\ }\href@noop {} {\emph {\bibinfo {title} {LEReC photocathode DC
  gun beam test results}}},\ \bibinfo {type} {Tech. Rep.}\ (\bibinfo
  {institution} {Brookhaven National Lab.(BNL), Upton, NY (United States)},\
  \bibinfo {year} {2018})\BibitemShut {NoStop}%
\bibitem [{\citenamefont {Nishimori}\ \emph {et~al.}(2013)\citenamefont
  {Nishimori}, \citenamefont {Nagai}, \citenamefont {Matsuba}, \citenamefont
  {Hajima}, \citenamefont {Yamamoto}, \citenamefont {Miyajima}, \citenamefont
  {Honda}, \citenamefont {Iijima}, \citenamefont {Kuriki},\ and\ \citenamefont
  {Kuwahara}}]{nishimori2013generation}%
  \BibitemOpen
  \bibfield  {author} {\bibinfo {author} {\bibfnamefont {N.}~\bibnamefont
  {Nishimori}}, \bibinfo {author} {\bibfnamefont {R.}~\bibnamefont {Nagai}},
  \bibinfo {author} {\bibfnamefont {S.}~\bibnamefont {Matsuba}}, \bibinfo
  {author} {\bibfnamefont {R.}~\bibnamefont {Hajima}}, \bibinfo {author}
  {\bibfnamefont {M.}~\bibnamefont {Yamamoto}}, \bibinfo {author}
  {\bibfnamefont {T.}~\bibnamefont {Miyajima}}, \bibinfo {author}
  {\bibfnamefont {Y.}~\bibnamefont {Honda}}, \bibinfo {author} {\bibfnamefont
  {H.}~\bibnamefont {Iijima}}, \bibinfo {author} {\bibfnamefont
  {M.}~\bibnamefont {Kuriki}},\ and\ \bibinfo {author} {\bibfnamefont
  {M.}~\bibnamefont {Kuwahara}},\ }\bibfield  {title} {\bibinfo {title}
  {Generation of a 500-kev electron beam from a high voltage photoemission
  gun},\ }\href@noop {} {\bibfield  {journal} {\bibinfo  {journal} {Applied
  Physics Letters}\ }\textbf {\bibinfo {volume} {102}},\ \bibinfo {pages}
  {234103} (\bibinfo {year} {2013})}\BibitemShut {NoStop}%
\bibitem [{\citenamefont {Hernandez-Garcia}\ \emph {et~al.}(2005)\citenamefont
  {Hernandez-Garcia}, \citenamefont {Siggins}, \citenamefont {Benson},
  \citenamefont {Bullard}, \citenamefont {Dylla}, \citenamefont {Jordan},
  \citenamefont {Murray}, \citenamefont {Neil}, \citenamefont {Shinn},\ and\
  \citenamefont {Walker}}]{hernandez2005high}%
  \BibitemOpen
  \bibfield  {author} {\bibinfo {author} {\bibfnamefont {C.}~\bibnamefont
  {Hernandez-Garcia}}, \bibinfo {author} {\bibfnamefont {T.}~\bibnamefont
  {Siggins}}, \bibinfo {author} {\bibfnamefont {S.}~\bibnamefont {Benson}},
  \bibinfo {author} {\bibfnamefont {D.}~\bibnamefont {Bullard}}, \bibinfo
  {author} {\bibfnamefont {H.}~\bibnamefont {Dylla}}, \bibinfo {author}
  {\bibfnamefont {K.}~\bibnamefont {Jordan}}, \bibinfo {author} {\bibfnamefont
  {C.}~\bibnamefont {Murray}}, \bibinfo {author} {\bibfnamefont
  {G.}~\bibnamefont {Neil}}, \bibinfo {author} {\bibfnamefont {M.}~\bibnamefont
  {Shinn}},\ and\ \bibinfo {author} {\bibfnamefont {R.}~\bibnamefont
  {Walker}},\ }\bibfield  {title} {\bibinfo {title} {A high average current dc
  gaas photocathode gun for erls and fels},\ }in\ \href@noop {} {\emph
  {\bibinfo {booktitle} {Proceedings of the 2005 Particle Accelerator
  Conference}}}\ (\bibinfo {organization} {IEEE},\ \bibinfo {year} {2005})\
  pp.\ \bibinfo {pages} {3117--3119}\BibitemShut {NoStop}%
\bibitem [{\citenamefont {Abbott}\ \emph {et~al.}(2016)\citenamefont {Abbott},
  \citenamefont {Adderley}, \citenamefont {Adeyemi}, \citenamefont {Aguilera},
  \citenamefont {Ali}, \citenamefont {Areti}, \citenamefont {Baylac},
  \citenamefont {Benesch}, \citenamefont {Bosson}, \citenamefont {Cade} \emph
  {et~al.}}]{abbott2016production}%
  \BibitemOpen
  \bibfield  {author} {\bibinfo {author} {\bibfnamefont {D.}~\bibnamefont
  {Abbott}}, \bibinfo {author} {\bibfnamefont {P.}~\bibnamefont {Adderley}},
  \bibinfo {author} {\bibfnamefont {A.}~\bibnamefont {Adeyemi}}, \bibinfo
  {author} {\bibfnamefont {P.}~\bibnamefont {Aguilera}}, \bibinfo {author}
  {\bibfnamefont {M.}~\bibnamefont {Ali}}, \bibinfo {author} {\bibfnamefont
  {H.}~\bibnamefont {Areti}}, \bibinfo {author} {\bibfnamefont
  {M.}~\bibnamefont {Baylac}}, \bibinfo {author} {\bibfnamefont
  {J.}~\bibnamefont {Benesch}}, \bibinfo {author} {\bibfnamefont
  {G.}~\bibnamefont {Bosson}}, \bibinfo {author} {\bibfnamefont
  {B.}~\bibnamefont {Cade}}, \emph {et~al.},\ }\bibfield  {title} {\bibinfo
  {title} {Production of highly polarized positrons using polarized electrons
  at mev energies},\ }\href@noop {} {\bibfield  {journal} {\bibinfo  {journal}
  {Physical review letters}\ }\textbf {\bibinfo {volume} {116}},\ \bibinfo
  {pages} {214801} (\bibinfo {year} {2016})}\BibitemShut {NoStop}%
\bibitem [{\citenamefont {Gu}\ \emph {et~al.}(2020)\citenamefont {Gu},
  \citenamefont {Altinbas}, \citenamefont {Badea}, \citenamefont {Bruno},
  \citenamefont {Cannizzo}, \citenamefont {Costanzo}, \citenamefont {Drees},
  \citenamefont {Fedotov}, \citenamefont {Fischer}, \citenamefont {Gassner}
  \emph {et~al.}}]{gu2020stable}%
  \BibitemOpen
  \bibfield  {author} {\bibinfo {author} {\bibfnamefont {X.}~\bibnamefont
  {Gu}}, \bibinfo {author} {\bibfnamefont {Z.}~\bibnamefont {Altinbas}},
  \bibinfo {author} {\bibfnamefont {S.}~\bibnamefont {Badea}}, \bibinfo
  {author} {\bibfnamefont {D.}~\bibnamefont {Bruno}}, \bibinfo {author}
  {\bibfnamefont {L.}~\bibnamefont {Cannizzo}}, \bibinfo {author}
  {\bibfnamefont {M.}~\bibnamefont {Costanzo}}, \bibinfo {author}
  {\bibfnamefont {A.}~\bibnamefont {Drees}}, \bibinfo {author} {\bibfnamefont
  {A.}~\bibnamefont {Fedotov}}, \bibinfo {author} {\bibfnamefont
  {W.}~\bibnamefont {Fischer}}, \bibinfo {author} {\bibfnamefont
  {D.}~\bibnamefont {Gassner}}, \emph {et~al.},\ }\bibfield  {title} {\bibinfo
  {title} {Stable operation of a high-voltage high-current dc photoemission gun
  for the bunched beam electron cooler in rhic},\ }\href@noop {} {\bibfield
  {journal} {\bibinfo  {journal} {Physical Review Accelerators and Beams}\
  }\textbf {\bibinfo {volume} {23}},\ \bibinfo {pages} {013401} (\bibinfo
  {year} {2020})}\BibitemShut {NoStop}%
\bibitem [{\citenamefont {Kilpatrick}(1957)}]{kilpatrick_criterion_1957}%
  \BibitemOpen
  \bibfield  {author} {\bibinfo {author} {\bibfnamefont {W.~D.}\ \bibnamefont
  {Kilpatrick}},\ }\bibfield  {title} {\bibinfo {title} {Criterion for {Vacuum}
  {Sparking} {Designed} to {Include} {Both} rf and dc},\ }\href
  {https://doi.org/10.1063/1.1715731} {\bibfield  {journal} {\bibinfo
  {journal} {Review of Scientific Instruments}\ }\textbf {\bibinfo {volume}
  {28}},\ \bibinfo {pages} {824} (\bibinfo {year} {1957})}\BibitemShut
  {NoStop}%
\bibitem [{\citenamefont {Rao}\ and\ \citenamefont
  {Dowell}(2013)}]{dowell_rao_book}%
  \BibitemOpen
  \bibinfo {editor} {\bibfnamefont {T.}~\bibnamefont {Rao}}\ and\ \bibinfo
  {editor} {\bibfnamefont {D.~H.}\ \bibnamefont {Dowell}},\ eds.,\ \href
  {https://arxiv.org/ftp/arxiv/papers/1403/1403.7539.pdf} {\emph {\bibinfo
  {title} {An Engineering Guide to Photoinjectors}}}\ (\bibinfo  {publisher}
  {CreateSpace Independent Publishing Platform},\ \bibinfo {year}
  {2013})\BibitemShut {NoStop}%
\bibitem [{\citenamefont {Norem}\ \emph {et~al.}(2005)\citenamefont {Norem},
  \citenamefont {Insepov},\ and\ \citenamefont
  {Konkashbaev}}]{norem_triggers_2005}%
  \BibitemOpen
  \bibfield  {author} {\bibinfo {author} {\bibfnamefont {J.}~\bibnamefont
  {Norem}}, \bibinfo {author} {\bibfnamefont {Z.}~\bibnamefont {Insepov}},\
  and\ \bibinfo {author} {\bibfnamefont {I.}~\bibnamefont {Konkashbaev}},\
  }\bibfield  {title} {\bibinfo {title} {Triggers for {RF} breakdown},\ }\href
  {https://doi.org/10.1016/j.nima.2004.07.289} {\bibfield  {journal} {\bibinfo
  {journal} {Nuclear Instruments and Methods in Physics Research Section A:
  Accelerators, Spectrometers, Detectors and Associated Equipment}\ }\textbf
  {\bibinfo {volume} {537}},\ \bibinfo {pages} {510} (\bibinfo {year}
  {2005})}\BibitemShut {NoStop}%
\bibitem [{\citenamefont {Tan}\ \emph {et~al.}(2022)\citenamefont {Tan},
  \citenamefont {Antipov}, \citenamefont {Doran}, \citenamefont {Ha},
  \citenamefont {Jing}, \citenamefont {Knight}, \citenamefont {Kuzikov},
  \citenamefont {Liu}, \citenamefont {Lu}, \citenamefont {Piot}, \citenamefont
  {Power}, \citenamefont {Shao}, \citenamefont {Whiteford},\ and\ \citenamefont
  {Wisniewski}}]{AWA_TW}%
  \BibitemOpen
  \bibfield  {author} {\bibinfo {author} {\bibfnamefont {W.~H.}\ \bibnamefont
  {Tan}}, \bibinfo {author} {\bibfnamefont {S.}~\bibnamefont {Antipov}},
  \bibinfo {author} {\bibfnamefont {D.~S.}\ \bibnamefont {Doran}}, \bibinfo
  {author} {\bibfnamefont {G.}~\bibnamefont {Ha}}, \bibinfo {author}
  {\bibfnamefont {C.}~\bibnamefont {Jing}}, \bibinfo {author} {\bibfnamefont
  {E.}~\bibnamefont {Knight}}, \bibinfo {author} {\bibfnamefont
  {S.}~\bibnamefont {Kuzikov}}, \bibinfo {author} {\bibfnamefont
  {W.}~\bibnamefont {Liu}}, \bibinfo {author} {\bibfnamefont {X.}~\bibnamefont
  {Lu}}, \bibinfo {author} {\bibfnamefont {P.}~\bibnamefont {Piot}}, \bibinfo
  {author} {\bibfnamefont {J.~G.}\ \bibnamefont {Power}}, \bibinfo {author}
  {\bibfnamefont {J.}~\bibnamefont {Shao}}, \bibinfo {author} {\bibfnamefont
  {C.}~\bibnamefont {Whiteford}},\ and\ \bibinfo {author} {\bibfnamefont
  {E.~E.}\ \bibnamefont {Wisniewski}},\ }\href
  {https://doi.org/10.48550/ARXIV.2203.11598} {\bibinfo {title} {Demonstration
  of sub-gv/m accelerating field in a photoemission electron gun powered by
  nanosecond $x$-band radiofrequency pulses}} (\bibinfo {year}
  {2022})\BibitemShut {NoStop}%
\bibitem [{\citenamefont {Cahill}\ \emph
  {et~al.}(2018{\natexlab{a}})\citenamefont {Cahill}, \citenamefont
  {Rosenzweig}, \citenamefont {Dolgashev}, \citenamefont {Tantawi},\ and\
  \citenamefont {Weathersby}}]{Cahill_cryoxband_18}%
  \BibitemOpen
  \bibfield  {author} {\bibinfo {author} {\bibfnamefont {A.~D.}\ \bibnamefont
  {Cahill}}, \bibinfo {author} {\bibfnamefont {J.~B.}\ \bibnamefont
  {Rosenzweig}}, \bibinfo {author} {\bibfnamefont {V.~A.}\ \bibnamefont
  {Dolgashev}}, \bibinfo {author} {\bibfnamefont {S.~G.}\ \bibnamefont
  {Tantawi}},\ and\ \bibinfo {author} {\bibfnamefont {S.}~\bibnamefont
  {Weathersby}},\ }\bibfield  {title} {\bibinfo {title} {High gradient
  experiments with $x$-band cryogenic copper accelerating cavities},\ }\href
  {https://doi.org/10.1103/PhysRevAccelBeams.21.102002} {\bibfield  {journal}
  {\bibinfo  {journal} {Phys. Rev. Accel. Beams}\ }\textbf {\bibinfo {volume}
  {21}},\ \bibinfo {pages} {102002} (\bibinfo {year}
  {2018}{\natexlab{a}})}\BibitemShut {NoStop}%
\bibitem [{\citenamefont {Cahill}\ \emph
  {et~al.}(2018{\natexlab{b}})\citenamefont {Cahill}, \citenamefont
  {Rosenzweig}, \citenamefont {Dolgashev}, \citenamefont {Li}, \citenamefont
  {Tantawi},\ and\ \citenamefont {Weathersby}}]{Cahill_rfloss_18}%
  \BibitemOpen
  \bibfield  {author} {\bibinfo {author} {\bibfnamefont {A.~D.}\ \bibnamefont
  {Cahill}}, \bibinfo {author} {\bibfnamefont {J.~B.}\ \bibnamefont
  {Rosenzweig}}, \bibinfo {author} {\bibfnamefont {V.~A.}\ \bibnamefont
  {Dolgashev}}, \bibinfo {author} {\bibfnamefont {Z.}~\bibnamefont {Li}},
  \bibinfo {author} {\bibfnamefont {S.~G.}\ \bibnamefont {Tantawi}},\ and\
  \bibinfo {author} {\bibfnamefont {S.}~\bibnamefont {Weathersby}},\ }\bibfield
   {title} {\bibinfo {title} {rf losses in a high gradient cryogenic copper
  cavity},\ }\href {https://doi.org/10.1103/PhysRevAccelBeams.21.061301}
  {\bibfield  {journal} {\bibinfo  {journal} {Phys. Rev. Accel. Beams}\
  }\textbf {\bibinfo {volume} {21}},\ \bibinfo {pages} {061301} (\bibinfo
  {year} {2018}{\natexlab{b}})}\BibitemShut {NoStop}%
\bibitem [{\citenamefont {Rosenzweig}\ \emph {et~al.}(2019)\citenamefont
  {Rosenzweig}, \citenamefont {Cahill}, \citenamefont {Dolgashev},
  \citenamefont {Emma}, \citenamefont {Fukasawa}, \citenamefont {Li},
  \citenamefont {Limborg}, \citenamefont {Maxson}, \citenamefont {Musumeci},
  \citenamefont {Nause}, \citenamefont {Pakter}, \citenamefont {Pompili},
  \citenamefont {Roussel}, \citenamefont {Spataro},\ and\ \citenamefont
  {Tantawi}}]{Rosenzweig_cryogun_19}%
  \BibitemOpen
  \bibfield  {author} {\bibinfo {author} {\bibfnamefont {J.~B.}\ \bibnamefont
  {Rosenzweig}}, \bibinfo {author} {\bibfnamefont {A.}~\bibnamefont {Cahill}},
  \bibinfo {author} {\bibfnamefont {V.}~\bibnamefont {Dolgashev}}, \bibinfo
  {author} {\bibfnamefont {C.}~\bibnamefont {Emma}}, \bibinfo {author}
  {\bibfnamefont {A.}~\bibnamefont {Fukasawa}}, \bibinfo {author}
  {\bibfnamefont {R.}~\bibnamefont {Li}}, \bibinfo {author} {\bibfnamefont
  {C.}~\bibnamefont {Limborg}}, \bibinfo {author} {\bibfnamefont
  {J.}~\bibnamefont {Maxson}}, \bibinfo {author} {\bibfnamefont
  {P.}~\bibnamefont {Musumeci}}, \bibinfo {author} {\bibfnamefont
  {A.}~\bibnamefont {Nause}}, \bibinfo {author} {\bibfnamefont
  {R.}~\bibnamefont {Pakter}}, \bibinfo {author} {\bibfnamefont
  {R.}~\bibnamefont {Pompili}}, \bibinfo {author} {\bibfnamefont
  {R.}~\bibnamefont {Roussel}}, \bibinfo {author} {\bibfnamefont
  {B.}~\bibnamefont {Spataro}},\ and\ \bibinfo {author} {\bibfnamefont
  {S.}~\bibnamefont {Tantawi}},\ }\bibfield  {title} {\bibinfo {title} {Next
  generation high brightness electron beams from ultrahigh field cryogenic rf
  photocathode sources},\ }\href
  {https://doi.org/10.1103/PhysRevAccelBeams.22.023403} {\bibfield  {journal}
  {\bibinfo  {journal} {Phys. Rev. Accel. Beams}\ }\textbf {\bibinfo {volume}
  {22}},\ \bibinfo {pages} {023403} (\bibinfo {year} {2019})}\BibitemShut
  {NoStop}%
\bibitem [{\citenamefont {Sannibale}\ \emph {et~al.}(2012)\citenamefont
  {Sannibale}, \citenamefont {Filippetto}, \citenamefont {Papadopoulos},
  \citenamefont {Staples}, \citenamefont {Wells}, \citenamefont {Bailey},
  \citenamefont {Baptiste}, \citenamefont {Corlett}, \citenamefont {Cork},
  \citenamefont {De~Santis}, \citenamefont {Dimaggio}, \citenamefont
  {Doolittle}, \citenamefont {Doyle}, \citenamefont {Feng}, \citenamefont
  {Garcia~Quintas}, \citenamefont {Huang}, \citenamefont {Huang}, \citenamefont
  {Kramasz}, \citenamefont {Kwiatkowski}, \citenamefont {Lellinger},
  \citenamefont {Moroz}, \citenamefont {Norum}, \citenamefont {Padmore},
  \citenamefont {Pappas}, \citenamefont {Portmann}, \citenamefont {Vecchione},
  \citenamefont {Vinco}, \citenamefont {Zolotorev},\ and\ \citenamefont
  {Zucca}}]{APEXgun}%
  \BibitemOpen
  \bibfield  {author} {\bibinfo {author} {\bibfnamefont {F.}~\bibnamefont
  {Sannibale}}, \bibinfo {author} {\bibfnamefont {D.}~\bibnamefont
  {Filippetto}}, \bibinfo {author} {\bibfnamefont {C.~F.}\ \bibnamefont
  {Papadopoulos}}, \bibinfo {author} {\bibfnamefont {J.}~\bibnamefont
  {Staples}}, \bibinfo {author} {\bibfnamefont {R.}~\bibnamefont {Wells}},
  \bibinfo {author} {\bibfnamefont {B.}~\bibnamefont {Bailey}}, \bibinfo
  {author} {\bibfnamefont {K.}~\bibnamefont {Baptiste}}, \bibinfo {author}
  {\bibfnamefont {J.}~\bibnamefont {Corlett}}, \bibinfo {author} {\bibfnamefont
  {C.}~\bibnamefont {Cork}}, \bibinfo {author} {\bibfnamefont {S.}~\bibnamefont
  {De~Santis}}, \bibinfo {author} {\bibfnamefont {S.}~\bibnamefont {Dimaggio}},
  \bibinfo {author} {\bibfnamefont {L.}~\bibnamefont {Doolittle}}, \bibinfo
  {author} {\bibfnamefont {J.}~\bibnamefont {Doyle}}, \bibinfo {author}
  {\bibfnamefont {J.}~\bibnamefont {Feng}}, \bibinfo {author} {\bibfnamefont
  {D.}~\bibnamefont {Garcia~Quintas}}, \bibinfo {author} {\bibfnamefont
  {G.}~\bibnamefont {Huang}}, \bibinfo {author} {\bibfnamefont
  {H.}~\bibnamefont {Huang}}, \bibinfo {author} {\bibfnamefont
  {T.}~\bibnamefont {Kramasz}}, \bibinfo {author} {\bibfnamefont
  {S.}~\bibnamefont {Kwiatkowski}}, \bibinfo {author} {\bibfnamefont
  {R.}~\bibnamefont {Lellinger}}, \bibinfo {author} {\bibfnamefont
  {V.}~\bibnamefont {Moroz}}, \bibinfo {author} {\bibfnamefont {W.~E.}\
  \bibnamefont {Norum}}, \bibinfo {author} {\bibfnamefont {H.}~\bibnamefont
  {Padmore}}, \bibinfo {author} {\bibfnamefont {C.}~\bibnamefont {Pappas}},
  \bibinfo {author} {\bibfnamefont {G.}~\bibnamefont {Portmann}}, \bibinfo
  {author} {\bibfnamefont {T.}~\bibnamefont {Vecchione}}, \bibinfo {author}
  {\bibfnamefont {M.}~\bibnamefont {Vinco}}, \bibinfo {author} {\bibfnamefont
  {M.}~\bibnamefont {Zolotorev}},\ and\ \bibinfo {author} {\bibfnamefont
  {F.}~\bibnamefont {Zucca}},\ }\bibfield  {title} {\bibinfo {title} {Advanced
  photoinjector experiment photogun commissioning results},\ }\href
  {https://doi.org/10.1103/PhysRevSTAB.15.103501} {\bibfield  {journal}
  {\bibinfo  {journal} {Physical Review Special Topics - Accelerators and
  Beams}\ }\textbf {\bibinfo {volume} {15}},\ \bibinfo {pages} {103501}
  (\bibinfo {year} {2012})}\BibitemShut {NoStop}%
\bibitem [{\citenamefont {Filippetto}\ and\ \citenamefont
  {Qian}(2016)}]{Filippetto2016}%
  \BibitemOpen
  \bibfield  {author} {\bibinfo {author} {\bibfnamefont {D.}~\bibnamefont
  {Filippetto}}\ and\ \bibinfo {author} {\bibfnamefont {H.}~\bibnamefont
  {Qian}},\ }\bibfield  {title} {\bibinfo {title} {{Design of a high-flux
  instrument for ultrafast electron diffraction and microscopy}},\ }\bibfield
  {journal} {\bibinfo  {journal} {Journal of Physics B: Atomic, Molecular and
  Optical Physics}\ }\textbf {\bibinfo {volume} {49}},\ \href
  {https://doi.org/10.1088/0953-4075/49/10/104003}
  {10.1088/0953-4075/49/10/104003} (\bibinfo {year} {2016})\BibitemShut
  {NoStop}%
\bibitem [{\citenamefont {Schmerge}\ \emph {et~al.}(2014)\citenamefont
  {Schmerge}, \citenamefont {Brachmann}, \citenamefont {Dowell}, \citenamefont
  {Fry}, \citenamefont {Li}, \citenamefont {Li}, \citenamefont {Raubenheimer},
  \citenamefont {Vecchione}, \citenamefont {Zhou}, \citenamefont {Bartnik},
  \citenamefont {Bazarov}, \citenamefont {Dunham}, \citenamefont {Gulliford},
  \citenamefont {Mayes}, \citenamefont {Filippetto}, \citenamefont {Huang},
  \citenamefont {Papadopoulos}, \citenamefont {Portmann}, \citenamefont
  {Qiang}, \citenamefont {Sannibale}, \citenamefont {Virostek}, \citenamefont
  {Wells}, \citenamefont {Lunin}, \citenamefont {Solyak},\ and\ \citenamefont
  {Vivoli}}]{lclsiigun2014}%
  \BibitemOpen
  \bibfield  {author} {\bibinfo {author} {\bibfnamefont {J.}~\bibnamefont
  {Schmerge}}, \bibinfo {author} {\bibfnamefont {A.}~\bibnamefont {Brachmann}},
  \bibinfo {author} {\bibfnamefont {D.}~\bibnamefont {Dowell}}, \bibinfo
  {author} {\bibfnamefont {A.}~\bibnamefont {Fry}}, \bibinfo {author}
  {\bibfnamefont {R.}~\bibnamefont {Li}}, \bibinfo {author} {\bibfnamefont
  {Z.}~\bibnamefont {Li}}, \bibinfo {author} {\bibfnamefont {T.}~\bibnamefont
  {Raubenheimer}}, \bibinfo {author} {\bibfnamefont {T.}~\bibnamefont
  {Vecchione}}, \bibinfo {author} {\bibfnamefont {F.}~\bibnamefont {Zhou}},
  \bibinfo {author} {\bibfnamefont {A.}~\bibnamefont {Bartnik}}, \bibinfo
  {author} {\bibfnamefont {I.}~\bibnamefont {Bazarov}}, \bibinfo {author}
  {\bibfnamefont {B.}~\bibnamefont {Dunham}}, \bibinfo {author} {\bibfnamefont
  {C.}~\bibnamefont {Gulliford}}, \bibinfo {author} {\bibfnamefont
  {C.}~\bibnamefont {Mayes}}, \bibinfo {author} {\bibfnamefont
  {D.}~\bibnamefont {Filippetto}}, \bibinfo {author} {\bibfnamefont
  {R.}~\bibnamefont {Huang}}, \bibinfo {author} {\bibfnamefont {C.~F.}\
  \bibnamefont {Papadopoulos}}, \bibinfo {author} {\bibfnamefont
  {G.}~\bibnamefont {Portmann}}, \bibinfo {author} {\bibfnamefont
  {J.}~\bibnamefont {Qiang}}, \bibinfo {author} {\bibfnamefont
  {F.}~\bibnamefont {Sannibale}}, \bibinfo {author} {\bibfnamefont
  {S.}~\bibnamefont {Virostek}}, \bibinfo {author} {\bibfnamefont
  {R.}~\bibnamefont {Wells}}, \bibinfo {author} {\bibfnamefont
  {A.}~\bibnamefont {Lunin}}, \bibinfo {author} {\bibfnamefont
  {N.}~\bibnamefont {Solyak}},\ and\ \bibinfo {author} {\bibfnamefont
  {A.}~\bibnamefont {Vivoli}},\ }\bibfield  {title} {\bibinfo {title} {The
  lcls-ii injector design},\ }in\ \href@noop {} {\emph {\bibinfo {booktitle}
  {Proceedings of FEL2014}}}\ (\bibinfo {address} {Basel, Switzerland},\
  \bibinfo {year} {2014})\BibitemShut {NoStop}%
\bibitem [{\citenamefont {Filippetto}\ \emph {et~al.}(2014)\citenamefont
  {Filippetto}, \citenamefont {Musumeci}, \citenamefont {Zolotorev},\ and\
  \citenamefont {Stupakov}}]{MaxCurrent}%
  \BibitemOpen
  \bibfield  {author} {\bibinfo {author} {\bibfnamefont {D.}~\bibnamefont
  {Filippetto}}, \bibinfo {author} {\bibfnamefont {P.}~\bibnamefont
  {Musumeci}}, \bibinfo {author} {\bibfnamefont {M.}~\bibnamefont
  {Zolotorev}},\ and\ \bibinfo {author} {\bibfnamefont {G.}~\bibnamefont
  {Stupakov}},\ }\bibfield  {title} {\bibinfo {title} {Maximum current density
  and beam brightness achievable by laser-driven electron sources},\ }\bibfield
   {journal} {\bibinfo  {journal} {Physical Review Special Topics -
  Accelerators and Beams}\ }\textbf {\bibinfo {volume} {17}},\ \href
  {https://doi.org/10.1103/PhysRevSTAB.17.024201}
  {10.1103/PhysRevSTAB.17.024201} (\bibinfo {year} {2014})\BibitemShut
  {NoStop}%
\bibitem [{\citenamefont {Filippetto}\ \emph
  {et~al.}(2015{\natexlab{b}})\citenamefont {Filippetto}, \citenamefont
  {Qian},\ and\ \citenamefont {Sannibale}}]{filippetto_cesium_2015}%
  \BibitemOpen
  \bibfield  {author} {\bibinfo {author} {\bibfnamefont {D.}~\bibnamefont
  {Filippetto}}, \bibinfo {author} {\bibfnamefont {H.}~\bibnamefont {Qian}},\
  and\ \bibinfo {author} {\bibfnamefont {F.}~\bibnamefont {Sannibale}},\
  }\bibfield  {title} {\bibinfo {title} {Cesium telluride cathodes for the next
  generation of high-average current high-brightness photoinjectors},\ }\href
  {https://doi.org/10.1063/1.4927700} {\bibfield  {journal} {\bibinfo
  {journal} {Applied Physics Letters}\ }\textbf {\bibinfo {volume} {107}},\
  \bibinfo {pages} {042104} (\bibinfo {year} {2015}{\natexlab{b}})}\BibitemShut
  {NoStop}%
\bibitem [{\citenamefont {Qian}\ and\ \citenamefont
  {Vogel}(2019)}]{qian_cwgun_2019}%
  \BibitemOpen
  \bibfield  {author} {\bibinfo {author} {\bibfnamefont {H.}~\bibnamefont
  {Qian}}\ and\ \bibinfo {author} {\bibfnamefont {E.}~\bibnamefont {Vogel}},\
  }\bibfield  {title} {\bibinfo {title} {{Overview of CW RF Guns for Short
  Wavelength FELs}},\ }in\ \href {https://doi.org/10.18429/JACoW-FEL2019-WEA01}
  {\emph {\bibinfo {booktitle} {Proc. FEL'19}}},\ \bibinfo {series and number}
  {\bibinfo {series} {Free Electron Laser Conference}\ No.~\bibinfo {number}
  {39}}\ (\bibinfo  {publisher} {JACoW Publishing, Geneva, Switzerland},\
  \bibinfo {year} {2019})\ pp.\ \bibinfo {pages} {290--296},\ \bibinfo {note}
  {https://doi.org/10.18429/JACoW-FEL2019-WEA01}\BibitemShut {NoStop}%
\bibitem [{\citenamefont {Shu}\ \emph {et~al.}(2019)\citenamefont {Shu},
  \citenamefont {Chen}, \citenamefont {Lal}, \citenamefont {Qian},
  \citenamefont {Shaker},\ and\ \citenamefont {Stephan}}]{pitz_cwgun_2019}%
  \BibitemOpen
  \bibfield  {author} {\bibinfo {author} {\bibfnamefont {G.}~\bibnamefont
  {Shu}}, \bibinfo {author} {\bibfnamefont {Y.}~\bibnamefont {Chen}}, \bibinfo
  {author} {\bibfnamefont {S.}~\bibnamefont {Lal}}, \bibinfo {author}
  {\bibfnamefont {H.}~\bibnamefont {Qian}}, \bibinfo {author} {\bibfnamefont
  {H.}~\bibnamefont {Shaker}},\ and\ \bibinfo {author} {\bibfnamefont
  {F.}~\bibnamefont {Stephan}},\ }\bibfield  {title} {\bibinfo {title} {{FIRST}
  {DESIGN} {STUDIES} {OF} {A} {NC} {CW} {RF} {GUN} {FOR} {EUROPEAN} {XFEL}},\
  }in\ \href {https://doi.org/doi:10.18429/JACoW-IPAC2019-TUPRB010} {\emph
  {\bibinfo {booktitle} {Proc. 10th International Particle Accelerator
  Conference (IPAC'19), Melbourne, Australia, 19-24 May 2019}}},\ \bibinfo
  {series and number} {\bibinfo {series} {International Particle Accelerator
  Conference}\ No.~\bibinfo {number} {10}}\ (\bibinfo  {publisher} {JACoW
  Publishing},\ \bibinfo {address} {Geneva, Switzerland},\ \bibinfo {year}
  {2019})\ pp.\ \bibinfo {pages} {1698--1701},\ \bibinfo {note}
  {https://doi.org/10.18429/JACoW-IPAC2019-TUPRB010}\BibitemShut {NoStop}%
\bibitem [{\citenamefont {Sannibale}\ \emph {et~al.}(2017)\citenamefont
  {Sannibale}, \citenamefont {Filippetto}, \citenamefont {Johnson},
  \citenamefont {Li}, \citenamefont {Luo}, \citenamefont {Mitchell},
  \citenamefont {Staples}, \citenamefont {Virostek}, \citenamefont {Wells},\
  and\ \citenamefont {Byrd}}]{apexii2017}%
  \BibitemOpen
  \bibfield  {author} {\bibinfo {author} {\bibfnamefont {F.}~\bibnamefont
  {Sannibale}}, \bibinfo {author} {\bibfnamefont {D.}~\bibnamefont
  {Filippetto}}, \bibinfo {author} {\bibfnamefont {M.}~\bibnamefont {Johnson}},
  \bibinfo {author} {\bibfnamefont {D.}~\bibnamefont {Li}}, \bibinfo {author}
  {\bibfnamefont {T.}~\bibnamefont {Luo}}, \bibinfo {author} {\bibfnamefont
  {C.}~\bibnamefont {Mitchell}}, \bibinfo {author} {\bibfnamefont
  {J.}~\bibnamefont {Staples}}, \bibinfo {author} {\bibfnamefont
  {S.}~\bibnamefont {Virostek}}, \bibinfo {author} {\bibfnamefont
  {R.}~\bibnamefont {Wells}},\ and\ \bibinfo {author} {\bibfnamefont {J.~M.}\
  \bibnamefont {Byrd}},\ }\bibfield  {title} {\bibinfo {title} {Upgrade
  possibilities for continuous wave rf electron guns based on room-temperature
  very high frequency technology},\ }\href
  {https://doi.org/10.1103/PhysRevAccelBeams.20.113402} {\bibfield  {journal}
  {\bibinfo  {journal} {Phys. Rev. Accel. Beams}\ }\textbf {\bibinfo {volume}
  {20}},\ \bibinfo {pages} {113402} (\bibinfo {year} {2017})}\BibitemShut
  {NoStop}%
\bibitem [{\citenamefont {Xiang}\ \emph {et~al.}(2021)\citenamefont {Xiang}
  \emph {et~al.}}]{xiang2021review}%
  \BibitemOpen
  \bibfield  {author} {\bibinfo {author} {\bibfnamefont {R.}~\bibnamefont
  {Xiang}} \emph {et~al.},\ }\bibfield  {title} {\bibinfo {title} {Review of
  superconducting radio frequency gun},\ }in\ \href@noop {} {\emph {\bibinfo
  {booktitle} {12th Int. Particle Accelerator Conf.(IPAC’21), Campinas,
  Brazil}}}\ (\bibinfo {year} {2021})\BibitemShut {NoStop}%
\bibitem [{\citenamefont {Vogel}\ \emph {et~al.}(2019)\citenamefont {Vogel},
  \citenamefont {Barbanotti}, \citenamefont {Brinkmann}, \citenamefont
  {Buettner}, \citenamefont {Iversen}, \citenamefont {Jensch}, \citenamefont
  {Klinke}, \citenamefont {Kostin}, \citenamefont {M{\"o}ller}, \citenamefont
  {Muhs} \emph {et~al.}}]{vogel2019status}%
  \BibitemOpen
  \bibfield  {author} {\bibinfo {author} {\bibfnamefont {E.}~\bibnamefont
  {Vogel}}, \bibinfo {author} {\bibfnamefont {S.}~\bibnamefont {Barbanotti}},
  \bibinfo {author} {\bibfnamefont {A.}~\bibnamefont {Brinkmann}}, \bibinfo
  {author} {\bibfnamefont {T.}~\bibnamefont {Buettner}}, \bibinfo {author}
  {\bibfnamefont {J.}~\bibnamefont {Iversen}}, \bibinfo {author} {\bibfnamefont
  {K.}~\bibnamefont {Jensch}}, \bibinfo {author} {\bibfnamefont
  {D.}~\bibnamefont {Klinke}}, \bibinfo {author} {\bibfnamefont
  {D.}~\bibnamefont {Kostin}}, \bibinfo {author} {\bibfnamefont {W.-D.}\
  \bibnamefont {M{\"o}ller}}, \bibinfo {author} {\bibfnamefont
  {A.}~\bibnamefont {Muhs}}, \emph {et~al.},\ }\bibfield  {title} {\bibinfo
  {title} {Status of the all superconducting gun cavity at desy},\ }in\
  \href@noop {} {\emph {\bibinfo {booktitle} {19th Int. Conf. on RF
  Superconductivity (SRF’19), Dresden, Germany, 30 June-05 July 2019}}}\
  (\bibinfo {year} {2019})\ pp.\ \bibinfo {pages} {1087--1090}\BibitemShut
  {NoStop}%
\bibitem [{\citenamefont {Teichert}\ \emph {et~al.}(2021)\citenamefont
  {Teichert}, \citenamefont {Arnold}, \citenamefont {Ciovati}, \citenamefont
  {Deinert}, \citenamefont {Evtushenko}, \citenamefont {Justus}, \citenamefont
  {Klopf}, \citenamefont {Kneisel}, \citenamefont {Kovalev}, \citenamefont
  {Kuntzsch} \emph {et~al.}}]{teichert2021successful}%
  \BibitemOpen
  \bibfield  {author} {\bibinfo {author} {\bibfnamefont {J.}~\bibnamefont
  {Teichert}}, \bibinfo {author} {\bibfnamefont {A.}~\bibnamefont {Arnold}},
  \bibinfo {author} {\bibfnamefont {G.}~\bibnamefont {Ciovati}}, \bibinfo
  {author} {\bibfnamefont {J.-C.}\ \bibnamefont {Deinert}}, \bibinfo {author}
  {\bibfnamefont {P.}~\bibnamefont {Evtushenko}}, \bibinfo {author}
  {\bibfnamefont {M.}~\bibnamefont {Justus}}, \bibinfo {author} {\bibfnamefont
  {J.}~\bibnamefont {Klopf}}, \bibinfo {author} {\bibfnamefont
  {P.}~\bibnamefont {Kneisel}}, \bibinfo {author} {\bibfnamefont
  {S.}~\bibnamefont {Kovalev}}, \bibinfo {author} {\bibfnamefont
  {M.}~\bibnamefont {Kuntzsch}}, \emph {et~al.},\ }\bibfield  {title} {\bibinfo
  {title} {Successful user operation of a superconducting radio-frequency
  photoelectron gun with mg cathodes},\ }\href@noop {} {\bibfield  {journal}
  {\bibinfo  {journal} {Physical Review Accelerators and Beams}\ }\textbf
  {\bibinfo {volume} {24}},\ \bibinfo {pages} {033401} (\bibinfo {year}
  {2021})}\BibitemShut {NoStop}%
\bibitem [{\citenamefont {Teichert}\ \emph {et~al.}(2014)\citenamefont
  {Teichert}, \citenamefont {Arnold}, \citenamefont {B{\"u}ttig}, \citenamefont
  {Justus}, \citenamefont {Kamps}, \citenamefont {Lehnert}, \citenamefont {Lu},
  \citenamefont {Michel}, \citenamefont {Murcek}, \citenamefont {Rudolph} \emph
  {et~al.}}]{teichert2014free}%
  \BibitemOpen
  \bibfield  {author} {\bibinfo {author} {\bibfnamefont {J.}~\bibnamefont
  {Teichert}}, \bibinfo {author} {\bibfnamefont {A.}~\bibnamefont {Arnold}},
  \bibinfo {author} {\bibfnamefont {H.}~\bibnamefont {B{\"u}ttig}}, \bibinfo
  {author} {\bibfnamefont {M.}~\bibnamefont {Justus}}, \bibinfo {author}
  {\bibfnamefont {T.}~\bibnamefont {Kamps}}, \bibinfo {author} {\bibfnamefont
  {U.}~\bibnamefont {Lehnert}}, \bibinfo {author} {\bibfnamefont
  {P.}~\bibnamefont {Lu}}, \bibinfo {author} {\bibfnamefont {P.}~\bibnamefont
  {Michel}}, \bibinfo {author} {\bibfnamefont {P.}~\bibnamefont {Murcek}},
  \bibinfo {author} {\bibfnamefont {J.}~\bibnamefont {Rudolph}}, \emph
  {et~al.},\ }\bibfield  {title} {\bibinfo {title} {Free-electron laser
  operation with a superconducting radio-frequency photoinjector at elbe},\
  }\href@noop {} {\bibfield  {journal} {\bibinfo  {journal} {Nuclear
  Instruments and Methods in Physics Research Section A: Accelerators,
  Spectrometers, Detectors and Associated Equipment}\ }\textbf {\bibinfo
  {volume} {743}},\ \bibinfo {pages} {114} (\bibinfo {year}
  {2014})}\BibitemShut {NoStop}%
\bibitem [{\citenamefont {Xin}\ \emph {et~al.}(2016)\citenamefont {Xin},
  \citenamefont {Brutus}, \citenamefont {Belomestnykh}, \citenamefont
  {Ben-Zvi}, \citenamefont {Boulware}, \citenamefont {Grimm}, \citenamefont
  {Hayes}, \citenamefont {Litvinenko}, \citenamefont {Mernick}, \citenamefont
  {Narayan} \emph {et~al.}}]{xin2016design}%
  \BibitemOpen
  \bibfield  {author} {\bibinfo {author} {\bibfnamefont {T.}~\bibnamefont
  {Xin}}, \bibinfo {author} {\bibfnamefont {J.}~\bibnamefont {Brutus}},
  \bibinfo {author} {\bibfnamefont {S.~A.}\ \bibnamefont {Belomestnykh}},
  \bibinfo {author} {\bibfnamefont {I.}~\bibnamefont {Ben-Zvi}}, \bibinfo
  {author} {\bibfnamefont {C.}~\bibnamefont {Boulware}}, \bibinfo {author}
  {\bibfnamefont {T.}~\bibnamefont {Grimm}}, \bibinfo {author} {\bibfnamefont
  {T.}~\bibnamefont {Hayes}}, \bibinfo {author} {\bibfnamefont {V.~N.}\
  \bibnamefont {Litvinenko}}, \bibinfo {author} {\bibfnamefont
  {K.}~\bibnamefont {Mernick}}, \bibinfo {author} {\bibfnamefont
  {G.}~\bibnamefont {Narayan}}, \emph {et~al.},\ }\bibfield  {title} {\bibinfo
  {title} {Design of a high-bunch-charge 112-mhz superconducting rf
  photoemission electron source},\ }\href@noop {} {\bibfield  {journal}
  {\bibinfo  {journal} {Review of Scientific Instruments}\ }\textbf {\bibinfo
  {volume} {87}},\ \bibinfo {pages} {093303} (\bibinfo {year}
  {2016})}\BibitemShut {NoStop}%
\bibitem [{\citenamefont {Wang}\ \emph
  {et~al.}(2021{\natexlab{a}})\citenamefont {Wang}, \citenamefont {Litvinenko},
  \citenamefont {Pinayev}, \citenamefont {Gaowei}, \citenamefont {Skaritka},
  \citenamefont {Belomestnykh}, \citenamefont {Ben-Zvi}, \citenamefont
  {Brutus}, \citenamefont {Jing}, \citenamefont {Biswas}, \citenamefont {Ma},
  \citenamefont {Narayan}, \citenamefont {Petrushina}, \citenamefont {Rahman},
  \citenamefont {Xin}, \citenamefont {Rao}, \citenamefont {Severino},
  \citenamefont {Shih}, \citenamefont {Smith}, \citenamefont {Wang},\ and\
  \citenamefont {Wu}}]{Wang_2021sr}%
  \BibitemOpen
  \bibfield  {author} {\bibinfo {author} {\bibfnamefont {E.}~\bibnamefont
  {Wang}}, \bibinfo {author} {\bibfnamefont {V.~N.}\ \bibnamefont
  {Litvinenko}}, \bibinfo {author} {\bibfnamefont {I.}~\bibnamefont {Pinayev}},
  \bibinfo {author} {\bibfnamefont {M.}~\bibnamefont {Gaowei}}, \bibinfo
  {author} {\bibfnamefont {J.}~\bibnamefont {Skaritka}}, \bibinfo {author}
  {\bibfnamefont {S.}~\bibnamefont {Belomestnykh}}, \bibinfo {author}
  {\bibfnamefont {I.}~\bibnamefont {Ben-Zvi}}, \bibinfo {author} {\bibfnamefont
  {J.~C.}\ \bibnamefont {Brutus}}, \bibinfo {author} {\bibfnamefont
  {Y.}~\bibnamefont {Jing}}, \bibinfo {author} {\bibfnamefont {J.}~\bibnamefont
  {Biswas}}, \bibinfo {author} {\bibfnamefont {J.}~\bibnamefont {Ma}}, \bibinfo
  {author} {\bibfnamefont {G.}~\bibnamefont {Narayan}}, \bibinfo {author}
  {\bibfnamefont {I.}~\bibnamefont {Petrushina}}, \bibinfo {author}
  {\bibfnamefont {O.}~\bibnamefont {Rahman}}, \bibinfo {author} {\bibfnamefont
  {T.}~\bibnamefont {Xin}}, \bibinfo {author} {\bibfnamefont {T.}~\bibnamefont
  {Rao}}, \bibinfo {author} {\bibfnamefont {F.}~\bibnamefont {Severino}},
  \bibinfo {author} {\bibfnamefont {K.}~\bibnamefont {Shih}}, \bibinfo {author}
  {\bibfnamefont {K.}~\bibnamefont {Smith}}, \bibinfo {author} {\bibfnamefont
  {G.}~\bibnamefont {Wang}},\ and\ \bibinfo {author} {\bibfnamefont
  {Y.}~\bibnamefont {Wu}},\ }\bibfield  {title} {\bibinfo {title} {{Long
  lifetime of bialkali photocathodes operating in high gradient superconducting
  radio frequency gun}},\ }\href {https://doi.org/10.1038/s41598-021-83997-1}
  {\bibfield  {journal} {\bibinfo  {journal} {Scientific Reports}\ }\textbf
  {\bibinfo {volume} {11}},\ \bibinfo {pages} {1} (\bibinfo {year}
  {2021}{\natexlab{a}})}\BibitemShut {NoStop}%
\bibitem [{\citenamefont {Legg}\ \emph {et~al.}(2012)\citenamefont {Legg},
  \citenamefont {Bisognano}, \citenamefont {Bissen}, \citenamefont {Bosch},
  \citenamefont {Eisert}, \citenamefont {Fisher}, \citenamefont {Green},
  \citenamefont {Kleman}, \citenamefont {Kulpin}, \citenamefont {Lawler} \emph
  {et~al.}}]{legg2012status}%
  \BibitemOpen
  \bibfield  {author} {\bibinfo {author} {\bibfnamefont {R.}~\bibnamefont
  {Legg}}, \bibinfo {author} {\bibfnamefont {J.}~\bibnamefont {Bisognano}},
  \bibinfo {author} {\bibfnamefont {M.}~\bibnamefont {Bissen}}, \bibinfo
  {author} {\bibfnamefont {R.}~\bibnamefont {Bosch}}, \bibinfo {author}
  {\bibfnamefont {D.}~\bibnamefont {Eisert}}, \bibinfo {author} {\bibfnamefont
  {M.}~\bibnamefont {Fisher}}, \bibinfo {author} {\bibfnamefont
  {M.}~\bibnamefont {Green}}, \bibinfo {author} {\bibfnamefont
  {K.}~\bibnamefont {Kleman}}, \bibinfo {author} {\bibfnamefont
  {J.}~\bibnamefont {Kulpin}}, \bibinfo {author} {\bibfnamefont
  {J.}~\bibnamefont {Lawler}}, \emph {et~al.},\ }\bibfield  {title} {\bibinfo
  {title} {Status of the wisconsin srf gun},\ }\href@noop {} {\bibfield
  {journal} {\bibinfo  {journal} {Proceedings of IPAC}\ }\textbf {\bibinfo
  {volume} {661}} (\bibinfo {year} {2012})}\BibitemShut {NoStop}%
\bibitem [{\citenamefont {Fuchs}\ \emph {et~al.}(2022)\citenamefont {Fuchs},
  \citenamefont {Shadwick}, \citenamefont {Vafaei-Najafabadi}, \citenamefont
  {Thomas}, \citenamefont {Andonian}, \citenamefont {Büscher}, \citenamefont
  {Lehrach}, \citenamefont {Apsimon}, \citenamefont {Xia}, \citenamefont
  {Filippetto}, \citenamefont {Schroeder},\ and\ \citenamefont
  {Downer}}]{fuchs_snowmass_2022}%
  \BibitemOpen
  \bibfield  {author} {\bibinfo {author} {\bibfnamefont {M.}~\bibnamefont
  {Fuchs}}, \bibinfo {author} {\bibfnamefont {B.~A.}\ \bibnamefont {Shadwick}},
  \bibinfo {author} {\bibfnamefont {N.}~\bibnamefont {Vafaei-Najafabadi}},
  \bibinfo {author} {\bibfnamefont {A.~G.~R.}\ \bibnamefont {Thomas}}, \bibinfo
  {author} {\bibfnamefont {G.}~\bibnamefont {Andonian}}, \bibinfo {author}
  {\bibfnamefont {M.}~\bibnamefont {Büscher}}, \bibinfo {author}
  {\bibfnamefont {A.}~\bibnamefont {Lehrach}}, \bibinfo {author} {\bibfnamefont
  {O.}~\bibnamefont {Apsimon}}, \bibinfo {author} {\bibfnamefont
  {G.}~\bibnamefont {Xia}}, \bibinfo {author} {\bibfnamefont {D.}~\bibnamefont
  {Filippetto}}, \bibinfo {author} {\bibfnamefont {C.~B.}\ \bibnamefont
  {Schroeder}},\ and\ \bibinfo {author} {\bibfnamefont {M.~C.}\ \bibnamefont
  {Downer}},\ }\href {http://arxiv.org/abs/2203.08379} {\bibinfo {title}
  {Snowmass {Whitepaper} {AF6}: {Plasma}-{Based} {Particle} {Sources}}}
  (\bibinfo {year} {2022}),\ \bibinfo {note} {number: arXiv:2203.08379
  arXiv:2203.08379 [physics]}\BibitemShut {NoStop}%
\bibitem [{\citenamefont {Power}\ \emph {et~al.}(2005)\citenamefont {Power},
  \citenamefont {Wang}, \citenamefont {Conde}, \citenamefont {Gai},
  \citenamefont {Konecny}, \citenamefont {Liu},\ and\ \citenamefont
  {Yusof}}]{power-2005}%
  \BibitemOpen
  \bibfield  {author} {\bibinfo {author} {\bibfnamefont {J.}~\bibnamefont
  {Power}}, \bibinfo {author} {\bibfnamefont {H.}~\bibnamefont {Wang}},
  \bibinfo {author} {\bibfnamefont {M.}~\bibnamefont {Conde}}, \bibinfo
  {author} {\bibfnamefont {W.}~\bibnamefont {Gai}}, \bibinfo {author}
  {\bibfnamefont {R.}~\bibnamefont {Konecny}}, \bibinfo {author} {\bibfnamefont
  {W.}~\bibnamefont {Liu}},\ and\ \bibinfo {author} {\bibfnamefont
  {Z.}~\bibnamefont {Yusof}},\ }\bibfield  {title} {\bibinfo {title}
  {Transverse beam envelope measurements and the limitations of the 3-screen
  emittance method for space-charge dominated beams},\ }\href
  {https://doi.org/https://doi.org/10.1016/j.nima.2005.02.034} {\bibfield
  {journal} {\bibinfo  {journal} {Nuclear Instruments and Methods in Physics
  Research Section A: Accelerators, Spectrometers, Detectors and Associated
  Equipment}\ }\textbf {\bibinfo {volume} {546}},\ \bibinfo {pages} {345}
  (\bibinfo {year} {2005})}\BibitemShut {NoStop}%
\bibitem [{\citenamefont {Mironov}\ \emph {et~al.}(2016)\citenamefont
  {Mironov}, \citenamefont {Potemkin}, \citenamefont {Gacheva}, \citenamefont
  {Andrianov}, \citenamefont {Zelenogorskii}, \citenamefont {Krasilnikov},
  \citenamefont {Stephan},\ and\ \citenamefont {Khazanov}}]{Mironov:16}%
  \BibitemOpen
  \bibfield  {author} {\bibinfo {author} {\bibfnamefont {S.~Y.}\ \bibnamefont
  {Mironov}}, \bibinfo {author} {\bibfnamefont {A.~K.}\ \bibnamefont
  {Potemkin}}, \bibinfo {author} {\bibfnamefont {E.~I.}\ \bibnamefont
  {Gacheva}}, \bibinfo {author} {\bibfnamefont {A.~V.}\ \bibnamefont
  {Andrianov}}, \bibinfo {author} {\bibfnamefont {V.~V.}\ \bibnamefont
  {Zelenogorskii}}, \bibinfo {author} {\bibfnamefont {M.}~\bibnamefont
  {Krasilnikov}}, \bibinfo {author} {\bibfnamefont {F.}~\bibnamefont
  {Stephan}},\ and\ \bibinfo {author} {\bibfnamefont {E.~A.}\ \bibnamefont
  {Khazanov}},\ }\bibfield  {title} {\bibinfo {title} {Shaping of cylindrical
  and 3d ellipsoidal beams for electron photoinjector laser drivers},\ }\href
  {https://doi.org/10.1364/AO.55.001630} {\bibfield  {journal} {\bibinfo
  {journal} {Appl. Opt.}\ }\textbf {\bibinfo {volume} {55}},\ \bibinfo {pages}
  {1630} (\bibinfo {year} {2016})}\BibitemShut {NoStop}%
\bibitem [{\citenamefont {ILC-collaboration}(2022)}]{ILC-2022}%
  \BibitemOpen
  \bibfield  {author} {\bibinfo {author} {\bibnamefont {ILC-collaboration}},\
  }\href {https://doi.org/10.48550/ARXIV.2203.07622} {\bibinfo {title} {The
  international linear collider: Report to snowmass 2021}} (\bibinfo {year}
  {2022})\BibitemShut {NoStop}%
\bibitem [{\citenamefont {Adolphsen}\ \emph {et~al.}(2013)\citenamefont
  {Adolphsen}, \citenamefont {Barone}, \citenamefont {Barish}, \citenamefont
  {Buesser}, \citenamefont {Burrows}, \citenamefont {Carwardine}, \citenamefont
  {Clark}, \citenamefont {Durand}, \citenamefont {Dugan}, \citenamefont
  {Elsen}, \citenamefont {Enomoto}, \citenamefont {Foster}, \citenamefont
  {Fukuda}, \citenamefont {Gai}, \citenamefont {Gastal}, \citenamefont {Geng},
  \citenamefont {Ginsburg}, \citenamefont {Guiducci}, \citenamefont {Harrison},
  \citenamefont {Hayano}, \citenamefont {Kershaw}, \citenamefont {Kubo},
  \citenamefont {Kuchler}, \citenamefont {List}, \citenamefont {Liu},
  \citenamefont {Michizono}, \citenamefont {Nantista}, \citenamefont {Osborne},
  \citenamefont {Palmer}, \citenamefont {Paterson}, \citenamefont {Peterson},
  \citenamefont {Phinney}, \citenamefont {Pierini}, \citenamefont {Ross},
  \citenamefont {Rubin}, \citenamefont {Seryi}, \citenamefont {Sheppard},
  \citenamefont {Solyak}, \citenamefont {Stapnes}, \citenamefont {Tauchi},
  \citenamefont {Toge}, \citenamefont {Walker}, \citenamefont {Yamamoto},\ and\
  \citenamefont {Yokoya}}]{ilc_tdr}%
  \BibitemOpen
  \bibfield  {author} {\bibinfo {author} {\bibfnamefont {C.}~\bibnamefont
  {Adolphsen}}, \bibinfo {author} {\bibfnamefont {M.}~\bibnamefont {Barone}},
  \bibinfo {author} {\bibfnamefont {B.}~\bibnamefont {Barish}}, \bibinfo
  {author} {\bibfnamefont {K.}~\bibnamefont {Buesser}}, \bibinfo {author}
  {\bibfnamefont {P.}~\bibnamefont {Burrows}}, \bibinfo {author} {\bibfnamefont
  {J.}~\bibnamefont {Carwardine}}, \bibinfo {author} {\bibfnamefont
  {J.}~\bibnamefont {Clark}}, \bibinfo {author} {\bibfnamefont {H.~M.}\
  \bibnamefont {Durand}}, \bibinfo {author} {\bibfnamefont {G.}~\bibnamefont
  {Dugan}}, \bibinfo {author} {\bibfnamefont {E.}~\bibnamefont {Elsen}},
  \bibinfo {author} {\bibfnamefont {A.}~\bibnamefont {Enomoto}}, \bibinfo
  {author} {\bibfnamefont {B.}~\bibnamefont {Foster}}, \bibinfo {author}
  {\bibfnamefont {S.}~\bibnamefont {Fukuda}}, \bibinfo {author} {\bibfnamefont
  {W.}~\bibnamefont {Gai}}, \bibinfo {author} {\bibfnamefont {M.}~\bibnamefont
  {Gastal}}, \bibinfo {author} {\bibfnamefont {R.}~\bibnamefont {Geng}},
  \bibinfo {author} {\bibfnamefont {C.}~\bibnamefont {Ginsburg}}, \bibinfo
  {author} {\bibfnamefont {S.}~\bibnamefont {Guiducci}}, \bibinfo {author}
  {\bibfnamefont {M.}~\bibnamefont {Harrison}}, \bibinfo {author}
  {\bibfnamefont {H.}~\bibnamefont {Hayano}}, \bibinfo {author} {\bibfnamefont
  {K.}~\bibnamefont {Kershaw}}, \bibinfo {author} {\bibfnamefont
  {K.}~\bibnamefont {Kubo}}, \bibinfo {author} {\bibfnamefont {V.}~\bibnamefont
  {Kuchler}}, \bibinfo {author} {\bibfnamefont {B.}~\bibnamefont {List}},
  \bibinfo {author} {\bibfnamefont {W.}~\bibnamefont {Liu}}, \bibinfo {author}
  {\bibfnamefont {S.}~\bibnamefont {Michizono}}, \bibinfo {author}
  {\bibfnamefont {C.}~\bibnamefont {Nantista}}, \bibinfo {author}
  {\bibfnamefont {J.}~\bibnamefont {Osborne}}, \bibinfo {author} {\bibfnamefont
  {M.}~\bibnamefont {Palmer}}, \bibinfo {author} {\bibfnamefont {J.~M.}\
  \bibnamefont {Paterson}}, \bibinfo {author} {\bibfnamefont {T.}~\bibnamefont
  {Peterson}}, \bibinfo {author} {\bibfnamefont {N.}~\bibnamefont {Phinney}},
  \bibinfo {author} {\bibfnamefont {P.}~\bibnamefont {Pierini}}, \bibinfo
  {author} {\bibfnamefont {M.}~\bibnamefont {Ross}}, \bibinfo {author}
  {\bibfnamefont {D.}~\bibnamefont {Rubin}}, \bibinfo {author} {\bibfnamefont
  {A.}~\bibnamefont {Seryi}}, \bibinfo {author} {\bibfnamefont
  {J.}~\bibnamefont {Sheppard}}, \bibinfo {author} {\bibfnamefont
  {N.}~\bibnamefont {Solyak}}, \bibinfo {author} {\bibfnamefont
  {S.}~\bibnamefont {Stapnes}}, \bibinfo {author} {\bibfnamefont
  {T.}~\bibnamefont {Tauchi}}, \bibinfo {author} {\bibfnamefont
  {N.}~\bibnamefont {Toge}}, \bibinfo {author} {\bibfnamefont {N.}~\bibnamefont
  {Walker}}, \bibinfo {author} {\bibfnamefont {A.}~\bibnamefont {Yamamoto}},\
  and\ \bibinfo {author} {\bibfnamefont {K.}~\bibnamefont {Yokoya}},\ }\href
  {https://doi.org/10.48550/ARXIV.1306.6328} {\bibinfo {title} {The
  international linear collider technical design report - volume 3.ii:
  Accelerator baseline design}} (\bibinfo {year} {2013})\BibitemShut {NoStop}%
\bibitem [{\citenamefont {Xu}\ \emph {et~al.}(2022)\citenamefont {Xu},
  \citenamefont {Kuriki}, \citenamefont {Piot},\ and\ \citenamefont
  {Power}}]{xu-2022a}%
  \BibitemOpen
  \bibfield  {author} {\bibinfo {author} {\bibfnamefont {T.}~\bibnamefont
  {Xu}}, \bibinfo {author} {\bibfnamefont {M.}~\bibnamefont {Kuriki}}, \bibinfo
  {author} {\bibfnamefont {P.}~\bibnamefont {Piot}},\ and\ \bibinfo {author}
  {\bibfnamefont {J.}~\bibnamefont {Power}},\ }\href
  {https://doi.org/10.48550/ARXIV.2205.03736} {\bibinfo {title} {Proposal for a
  damping-ring-free electron injector for future linear colliders}} (\bibinfo
  {year} {2022})\BibitemShut {NoStop}%
\bibitem [{\citenamefont {Kim}(2003)}]{PhysRevSTAB.6.104002}%
  \BibitemOpen
  \bibfield  {author} {\bibinfo {author} {\bibfnamefont {K.-J.}\ \bibnamefont
  {Kim}},\ }\bibfield  {title} {\bibinfo {title} {Round-to-flat transformation
  of angular-momentum-dominated beams},\ }\href
  {https://doi.org/10.1103/PhysRevSTAB.6.104002} {\bibfield  {journal}
  {\bibinfo  {journal} {Phys. Rev. ST Accel. Beams}\ }\textbf {\bibinfo
  {volume} {6}},\ \bibinfo {pages} {104002} (\bibinfo {year}
  {2003})}\BibitemShut {NoStop}%
\bibitem [{\citenamefont {Burov}\ \emph {et~al.}(2002)\citenamefont {Burov},
  \citenamefont {Nagaitsev},\ and\ \citenamefont
  {Derbenev}}]{PhysRevE.66.016503}%
  \BibitemOpen
  \bibfield  {author} {\bibinfo {author} {\bibfnamefont {A.}~\bibnamefont
  {Burov}}, \bibinfo {author} {\bibfnamefont {S.}~\bibnamefont {Nagaitsev}},\
  and\ \bibinfo {author} {\bibfnamefont {Y.}~\bibnamefont {Derbenev}},\
  }\bibfield  {title} {\bibinfo {title} {Circular modes, beam adapters, and
  their applications in beam optics},\ }\href
  {https://doi.org/10.1103/PhysRevE.66.016503} {\bibfield  {journal} {\bibinfo
  {journal} {Phys. Rev. E}\ }\textbf {\bibinfo {volume} {66}},\ \bibinfo
  {pages} {016503} (\bibinfo {year} {2002})}\BibitemShut {NoStop}%
\bibitem [{\citenamefont {Piot}\ \emph
  {et~al.}(2006{\natexlab{b}})\citenamefont {Piot}, \citenamefont {Sun},\ and\
  \citenamefont {Kim}}]{PhysRevSTAB.9.031001}%
  \BibitemOpen
  \bibfield  {author} {\bibinfo {author} {\bibfnamefont {P.}~\bibnamefont
  {Piot}}, \bibinfo {author} {\bibfnamefont {Y.-E.}\ \bibnamefont {Sun}},\ and\
  \bibinfo {author} {\bibfnamefont {K.-J.}\ \bibnamefont {Kim}},\ }\bibfield
  {title} {\bibinfo {title} {Photoinjector generation of a flat electron beam
  with transverse emittance ratio of 100},\ }\href
  {https://doi.org/10.1103/PhysRevSTAB.9.031001} {\bibfield  {journal}
  {\bibinfo  {journal} {Phys. Rev. ST Accel. Beams}\ }\textbf {\bibinfo
  {volume} {9}},\ \bibinfo {pages} {031001} (\bibinfo {year}
  {2006}{\natexlab{b}})}\BibitemShut {NoStop}%
\bibitem [{\citenamefont {Wang}\ \emph
  {et~al.}(2021{\natexlab{b}})\citenamefont {Wang}, \citenamefont {Peggs},
  \citenamefont {Ptitsyn}, \citenamefont {Willeke}, \citenamefont {Xu},\ and\
  \citenamefont {Hoffstaetter}}]{wang2021accelerator}%
  \BibitemOpen
  \bibfield  {author} {\bibinfo {author} {\bibfnamefont {E.}~\bibnamefont
  {Wang}}, \bibinfo {author} {\bibfnamefont {S.}~\bibnamefont {Peggs}},
  \bibinfo {author} {\bibfnamefont {V.}~\bibnamefont {Ptitsyn}}, \bibinfo
  {author} {\bibfnamefont {F.}~\bibnamefont {Willeke}}, \bibinfo {author}
  {\bibfnamefont {W.}~\bibnamefont {Xu}},\ and\ \bibinfo {author}
  {\bibfnamefont {G.}~\bibnamefont {Hoffstaetter}},\ }\bibfield  {title}
  {\bibinfo {title} {The accelerator design progress for eic strong hadron
  cooling},\ }in\ \href@noop {} {\emph {\bibinfo {booktitle} {12th Int.
  Particle Accelerator Conf.(IPAC’21), Campinas, SP, Brazil}}}\ (\bibinfo
  {year} {2021})\BibitemShut {NoStop}%
\bibitem [{\citenamefont {Cornacchia}\ and\ \citenamefont
  {Emma}(2002)}]{PhysRevSTAB.5.084001}%
  \BibitemOpen
  \bibfield  {author} {\bibinfo {author} {\bibfnamefont {M.}~\bibnamefont
  {Cornacchia}}\ and\ \bibinfo {author} {\bibfnamefont {P.}~\bibnamefont
  {Emma}},\ }\bibfield  {title} {\bibinfo {title} {Transverse to longitudinal
  emittance exchange},\ }\href {https://doi.org/10.1103/PhysRevSTAB.5.084001}
  {\bibfield  {journal} {\bibinfo  {journal} {Phys. Rev. ST Accel. Beams}\
  }\textbf {\bibinfo {volume} {5}},\ \bibinfo {pages} {084001} (\bibinfo {year}
  {2002})}\BibitemShut {NoStop}%
\bibitem [{\citenamefont {Lucas}\ \emph {et~al.}(2021)\citenamefont {Lucas},
  \citenamefont {Stragier}, \citenamefont {Mutsaers},\ and\ \citenamefont
  {Luiten}}]{LUCAS2021165651}%
  \BibitemOpen
  \bibfield  {author} {\bibinfo {author} {\bibfnamefont {T.}~\bibnamefont
  {Lucas}}, \bibinfo {author} {\bibfnamefont {X.}~\bibnamefont {Stragier}},
  \bibinfo {author} {\bibfnamefont {P.}~\bibnamefont {Mutsaers}},\ and\
  \bibinfo {author} {\bibfnamefont {O.}~\bibnamefont {Luiten}},\ }\bibfield
  {title} {\bibinfo {title} {Rf design of a compact, x-band travelling-wave rf
  photogun made from halves},\ }\href
  {https://doi.org/https://doi.org/10.1016/j.nima.2021.165651} {\bibfield
  {journal} {\bibinfo  {journal} {Nuclear Instruments and Methods in Physics
  Research Section A: Accelerators, Spectrometers, Detectors and Associated
  Equipment}\ }\textbf {\bibinfo {volume} {1013}},\ \bibinfo {pages} {165651}
  (\bibinfo {year} {2021})}\BibitemShut {NoStop}%
\bibitem [{\citenamefont {Filippetto}\ \emph
  {et~al.}(2015{\natexlab{c}})\citenamefont {Filippetto}, \citenamefont
  {Qian},\ and\ \citenamefont {Sannibale}}]{vhf_15}%
  \BibitemOpen
  \bibfield  {author} {\bibinfo {author} {\bibfnamefont {D.}~\bibnamefont
  {Filippetto}}, \bibinfo {author} {\bibfnamefont {H.}~\bibnamefont {Qian}},\
  and\ \bibinfo {author} {\bibfnamefont {F.}~\bibnamefont {Sannibale}},\
  }\bibfield  {title} {\bibinfo {title} {Cesium telluride cathodes for the next
  generation of high-average current high-brightness photoinjectors},\ }\href
  {https://doi.org/10.1063/1.4927700} {\bibfield  {journal} {\bibinfo
  {journal} {Applied Physics Letters}\ }\textbf {\bibinfo {volume} {107}},\
  \bibinfo {pages} {042104} (\bibinfo {year} {2015}{\natexlab{c}})},\ \Eprint
  {https://arxiv.org/abs/https://doi.org/10.1063/1.4927700}
  {https://doi.org/10.1063/1.4927700} \BibitemShut {NoStop}%
\bibitem [{\citenamefont {Huang}\ \emph {et~al.}(2015)\citenamefont {Huang},
  \citenamefont {Filippetto}, \citenamefont {Papadopoulos}, \citenamefont
  {Qian}, \citenamefont {Sannibale},\ and\ \citenamefont
  {Zolotorev}}]{PhysRevSTAB.18.013401}%
  \BibitemOpen
  \bibfield  {author} {\bibinfo {author} {\bibfnamefont {R.}~\bibnamefont
  {Huang}}, \bibinfo {author} {\bibfnamefont {D.}~\bibnamefont {Filippetto}},
  \bibinfo {author} {\bibfnamefont {C.~F.}\ \bibnamefont {Papadopoulos}},
  \bibinfo {author} {\bibfnamefont {H.}~\bibnamefont {Qian}}, \bibinfo {author}
  {\bibfnamefont {F.}~\bibnamefont {Sannibale}},\ and\ \bibinfo {author}
  {\bibfnamefont {M.}~\bibnamefont {Zolotorev}},\ }\bibfield  {title} {\bibinfo
  {title} {Dark current studies on a normal-conducting high-brightness
  very-high-frequency electron gun operating in continuous wave mode},\ }\href
  {https://doi.org/10.1103/PhysRevSTAB.18.013401} {\bibfield  {journal}
  {\bibinfo  {journal} {Phys. Rev. ST Accel. Beams}\ }\textbf {\bibinfo
  {volume} {18}},\ \bibinfo {pages} {013401} (\bibinfo {year}
  {2015})}\BibitemShut {NoStop}%
\bibitem [{\citenamefont {Xiang}\ \emph {et~al.}(2014)\citenamefont {Xiang},
  \citenamefont {Arnold}, \citenamefont {Kamps}, \citenamefont {Lu},
  \citenamefont {Michel}, \citenamefont {Murcek}, \citenamefont {Vennekate},
  \citenamefont {Staats},\ and\ \citenamefont
  {Teichert}}]{PhysRevSTAB.17.043401}%
  \BibitemOpen
  \bibfield  {author} {\bibinfo {author} {\bibfnamefont {R.}~\bibnamefont
  {Xiang}}, \bibinfo {author} {\bibfnamefont {A.}~\bibnamefont {Arnold}},
  \bibinfo {author} {\bibfnamefont {T.}~\bibnamefont {Kamps}}, \bibinfo
  {author} {\bibfnamefont {P.}~\bibnamefont {Lu}}, \bibinfo {author}
  {\bibfnamefont {P.}~\bibnamefont {Michel}}, \bibinfo {author} {\bibfnamefont
  {P.}~\bibnamefont {Murcek}}, \bibinfo {author} {\bibfnamefont
  {H.}~\bibnamefont {Vennekate}}, \bibinfo {author} {\bibfnamefont
  {G.}~\bibnamefont {Staats}},\ and\ \bibinfo {author} {\bibfnamefont
  {J.}~\bibnamefont {Teichert}},\ }\bibfield  {title} {\bibinfo {title}
  {Experimental studies of dark current in a superconducting rf
  photoinjector},\ }\href {https://doi.org/10.1103/PhysRevSTAB.17.043401}
  {\bibfield  {journal} {\bibinfo  {journal} {Phys. Rev. ST Accel. Beams}\
  }\textbf {\bibinfo {volume} {17}},\ \bibinfo {pages} {043401} (\bibinfo
  {year} {2014})}\BibitemShut {NoStop}%
\end{thebibliography}%

\end{document}